\documentclass[preprint,12pt,authoryear]{elsarticle}

\usepackage{amssymb}
\usepackage{amsmath}
\usepackage{hyperref}
\usepackage{caption}
\usepackage{subcaption}

\journal{Preprint}

\begin{document}

\begin{frontmatter}

\title{Hetero-modal learning and corruption-resistant hetero-modal inference for joint segmentation of white matter hyperintensities and ischaemic stroke lesions in MRI}

\author[inst1,inst2]{Jesse Phitidis}
\author[inst1]{William N. Whiteley}
\author[inst1,inst6]{Joanna M. Wardlaw}
\author[inst7]{Miguel O. Bernabeu}
\author[inst2]{Francesco Dalla Serra}
\author[inst1,inst6]{Maria del C. Vald\'{e}s Hern\'{a}ndez\corref{cor1}}
\ead{M.Valdes-Hernan@ed.ac.uk}
\cortext[cor1]{Corresponding author}

\affiliation[inst1]{organization={Centre for Clinical Brain Sciences, University of Edinburgh},
            addressline={49 Little France Crescent}, 
            city={Edinburgh},
            postcode={EH164SB}, 
            country={United Kingdom}}

\affiliation[inst2]{organization={Canon Medical Research Europe},
            addressline={Bonnington Bond, 2 Anderson Place}, 
            city={Edinburgh},
            postcode={EH65NP}, 
            country={United Kingdom}}


\affiliation[inst6]{organization={UK Dementia Research Institute, Centre at The University of Edinburgh},
            addressline={49 Little France Crescent}, 
            city={Edinburgh},
            postcode={EH164SB}, 
            country={United Kingdom}}

\affiliation[inst7]{organization={Usher Institute, University of Edinburgh},
            addressline={NINE, 9 Little France Road}, 
            city={Edinburgh},
            postcode={EH164UX}, 
            country={United Kingdom}}

\begin{abstract}

White matter hyperintensities (WMH) and ischaemic stroke lesions (ISL) are visually confounding, co-occurring pathologies that require large, diverse datasets for robust deep learning segmentation. However, assembling such datasets is hindered by cohort samples that lack reference segmentations from both features and complete sets of MRI structural sequences (i.e., ``modalities''). To maximise data utility, we investigate hetero-modal learning using a dataset of 206 vascular disease patients across four MRI sequences (T1-weighted, T2-weighted, fluid-attenuated inversion recovery, and diffusion-weighted imaging) with expert annotations of both WMH and ISL. We demonstrate that hetero-modal learning outperforms models trained using a single imaging modality in scenarios with substantial missing data, including a split where only 10\% of the training data contains all four modalities while the remainder is uni-modal, and a split relying on a single shared ``anchor'' modality with zero overlap between the remaining modalities. Furthermore, models trained under this second split successfully perform inference on unseen combinations of modalities. Yet, while standard hetero-modal networks handle missing sequences, clinical deployment introduces the additional challenge of silent data degradation---where modalities are present but severely corrupted. To bridge this gap, we introduce the Multimodal Attention Router (MMAR) block. Our experiments demonstrate that, when trained with a ``corruption augmentation'' strategy, the MMAR effectively dynamically weights the encoded features of each modality, maintaining strong performance during hetero-modal inference even in the presence of unflagged catastrophically corrupted modalities.

\end{abstract}




\end{frontmatter}
\section{Introduction} \label{ch5_sec:intro}

White matter hyperintensities (WMH) and ischaemic stroke lesions (ISL) are pathological features often visible on magnetic resonance imaging (MRI) scans of individuals with cerebral small vessel disease (SVD) \citep{duering2023neuroimaging}, a vascular disease affecting the small blood vessels. Their automatic segmentation and differentiation is inherently challenging due to their similar visual appearance in most MRI sequences: both lesion types present as hypointense on T1-weighted (T1w) MRI sequences and hyperintense on T2-weighted (T2w) MRI sequences, including the fluid-attenuated inversion recovery (FLAIR) sequence. Although diffusion-weighted images (DWI) can sometimes distinguish ISL from WMH, this is typically only effective when ISL is acute and, as apparent diffusion coefficients vary with age, body temperature, scanning settings, and tissue specific characteristics, this MRI modality is not always a reliable way to identify ISL \citep{edlow2017diagnosis}.

\vspace{8pt}
While visually differentiating these confounding lesions is difficult, it is a task well suited to deep learning, as neural networks excel at uncovering highly complex patterns. To achieve robust performance, however, these models must be trained on as much expertly annotated data as possible, and this data should ideally originate from a large variety of sources (i.e., imaging acquisition protocols and MRI scanners). Beyond a general scarcity of such data, model development is hindered by two main factors: 1) not all datasets with WMH or ISL annotations have them for both lesion types; and 2) clinical datasets rarely provide a complete, pristine set of MRI sequences: sequences are frequently missing, or present but severely degraded by artefacts. The first issue describes the problem of learning from partially labelled datasets, which has been previously explored \citep{phitidis2026comparative}. In this paper, we focus on the second issue, investigating the merits of hetero-modal learning, defined as learning from heterogeneous subsets of all unique MRI sequence types available throughout the training dataset.

\vspace{8pt}
Whereas in clinical and medical physics the image acquisition method defines the imaging modality (e.g., computerised tomography, MRI, positron-emission tomography), in the field of computer science, it is becoming more common to refer to each distinct sequence type, even within MRI, as a modality. In this manuscript we have adopted this definition, for compatibility with current literature in the field of medical image segmentation. 

\vspace{8pt}
A standard multi-modal segmentation model requires a fixed set of input modalities. A hetero-modal model, however, is designed to generate predictions from any available subset of imaging modalities. The seminal work in this area is HeMIS (Hetero-modal Medical Image Segmentation) by \cite{havaei2016hemis}, which proposed the use of independent encoders to embed each modality into a shared latent space. Subsequent work \citep{dorent2019hetero} incorporated variational autoencoders to reconstruct the missing modalities. More recently, convolutional neural networks \citep{lecun1989backpropagation} have been combined with vision transformers \citep{vaswani2017attention} for more flexible hetero-modal processing \citep{zhang2022mmformer, liu2023sfusion, zhu2025bridging, lan2025dmaf}.

\vspace{8pt}
Much of the literature focuses on hetero-modal \textit{inference}, but hetero-modal \textit{learning} is arguably the more important aspect of the framework. The primary advantage of training on a dataset where not all samples have a complete set of modalities is the ability to include a larger number of individual subjects during training. This allows a model to capture a greater proportion of the anatomical and pathological variation present in the underlying clinical population. In theory, this also allows models to leverage ``privileged information''---meaning that learning from combined modalities can ultimately improve the model's single-modality performance by encouraging richer, correlated feature representations. In practice, however, this depends heavily on the dataset and the specific clinical task, and modality-specialist baselines still frequently outperform hetero-modal methods when trained on the same set of subjects.

\vspace{8pt}
The recent literature has begun to explore hetero-modal learning. \cite{shi2024passion} simulated imbalanced, incomplete modality scenarios in the BraTS \citep{menze2014multimodal} and MyOPS \citep{li2023myops} datasets. \cite{lan2025dmaf} studied incomplete hetero-modal learning under varying missing rates using BraTS. \cite{wagner2025feasibility} investigated federated learning in seven datasets comprising seven MRI sequences and five pathologies. Using a multi-channel U-Net \citep{ronneberger2015u} with zero-imputation (passing missing modalities into the network as channels of zeros), they showed their model could perform segmentation on unseen modality combinations. In particular, they used the WMH segmentation challenge dataset \citep{kuijf2019standardized} for WMH segmentation (T1w and FLAIR), alongside the ATLAS v2.0 \citep{liew2022large} (T1w only) and ISLES \citep{hernandez2022isles} (T1w, T2w, FLAIR, and DWI) datasets for ISL segmentation. However, all pathologies were merged into a single ``lesion'' class. While the WMH Challenge, ATLAS v2.0, and ISLES datasets contain ground truth segmentations for WMH or ISL independently, no publicly available dataset currently provides joint ground truth annotations for \textit{both} pathologies. This lack of joint labels restricts a model's ability to accurately penalise false negatives when these visually similar lesions co-occur. Furthermore, mild SVD-related ISL (i.e., lacunar strokes) are under-represented in the ATLAS and ISLES cohorts. To address this gap, we leverage a privately held dataset containing expertly annotated ground truth segmentations for both WMH and ISL, with four distinct MRI sequences available (T1w, T2w, FLAIR, and DWI). 

\vspace{8pt}
While utilising this unique dataset allows us to rigorously evaluate hetero-modal learning for joint segmentation of WMH and ISL, real-world clinical deployment presents an additional hurdle: silent data corruption. A limitation of current hetero-modal approaches is the assumption that missing modalities are always correctly identified by the system. In automated image processing workflows, this is frequently not the case. MRI files might be filled with zeros or other constant voxel values, or a scan might be corrupted by artefacts severe enough to make it unusable. To overcome this, we introduce an additional layer of extreme data augmentation which we refer to as ``corruption augmentation''. With this corruption augmentation in mind, we develop the Multimodal Attention Router (MMAR) block. While we do not comprehensively benchmark our MMAR-based model against the entirety of the hetero-modal literature, we present it as a proof-of-concept for a practical architectural solution designed to train corruption-robustness directly into the hetero-modal segmentation pipeline without diluting the representational capacity of the modality-specific encoders.

\vspace{8pt}
This paper makes the following contributions:

\begin{enumerate}
    \item We establish the first benchmark for joint hetero-modal segmentation of co-occurring WMH and ISL by leveraging a dataset of patients with subacute and chronic ISL and various degrees of vascular pathology, with ground truth annotations for both pathologies, and four MRI sequences.
    \item We show that hetero-modal learning can train significantly better models than fixed-modality baselines under a training data distribution containing only a small (10\%) full-modality subset, with the remainder being strictly uni-modal.
    \item We demonstrate that hetero-modal learning can produce models capable of performing inference on unseen combinations of modalities.
    \item To address the vulnerability of hetero-modal networks to silent data degradation, we introduce the Multimodal Attention Router (MMAR) block, an architectural solution designed to for dynamic feature routing, ensuring robustness against corrupted inputs.
    \item We show that when trained with a ``corruption augmentation'' strategy, our MMAR-based model maintains strong performance in the face of catastrophic noise, severe bias field corruption, and modalities incorrectly identified as being present.
\end{enumerate}

\section{Experimental setup}

\subsection{Dataset and processing}

We performed our experiments on the Mild Stroke Study 3 dataset \citep{clancy2021rationale}---a prospective observational cohort study of patients presenting with clinically evident stroke syndromes, which was granted ethical approval by Southeast Scotland Regional Ethics Committee (ref. 18/SS/0044).
T1w (3D protocol, $1.0\times1.0\times1.0$ mm voxels), T2w (3D protocol, $0.9\times0.9\times0.9$ mm voxels), FLAIR (3D protocol, $1.0\times1.0\times1.0$ mm voxels), and DWI (2D protocol, $2.0\times2.0\times2.0$ mm voxels) MRI scans were acquired using a 3 Tesla Siemens Prisma MRI scanner.
An expert manually delineated ISL masks, and WMH masks were produced via a semi-automatic approach and subsequently corrected by an expert \citep{valdes2023step}.
We selected the subset of 206 patients (mean age $\pm$ std = 66 $\pm$ 11; 33\% female; 69\% hypertensive; 21\% diabetics) with a complete set of MRI sequences available, in addition to WMH and ISL segmentation masks. Stroke sub-types included cortical and lacunar strokes affecting several locations including: middle cerebral artery, anterior cerebral artery, brain stem, cerebellum, internal and external capsules, lentiform nucleus, and the thalamus.

\begin{figure}[htbp]
     \centering
     \begin{subfigure}[b]{0.32\textwidth}
         \centering
         \includegraphics[width=\textwidth]{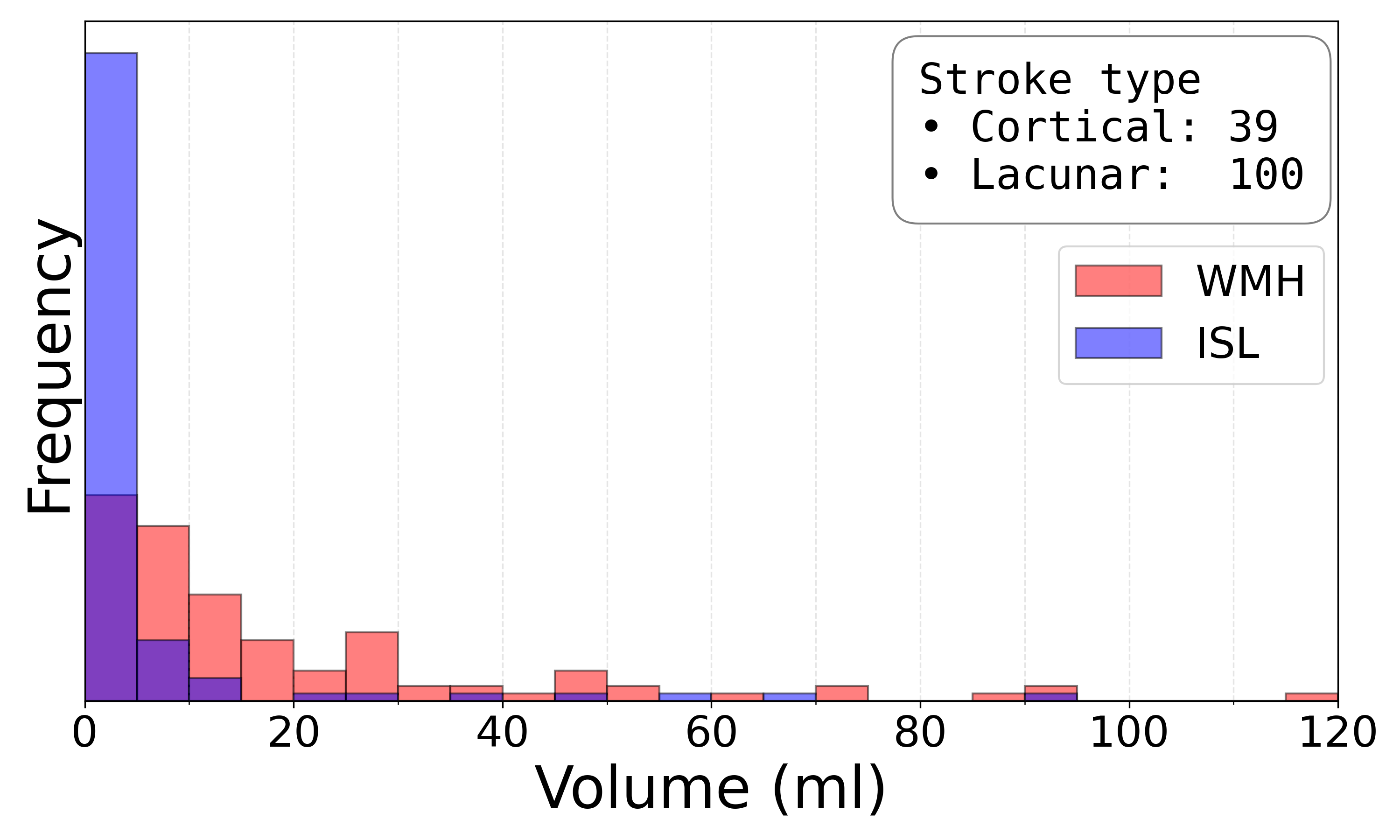}
     \end{subfigure}
     \hfill
     \begin{subfigure}[b]{0.32\textwidth}
         \centering
         \includegraphics[width=\textwidth]{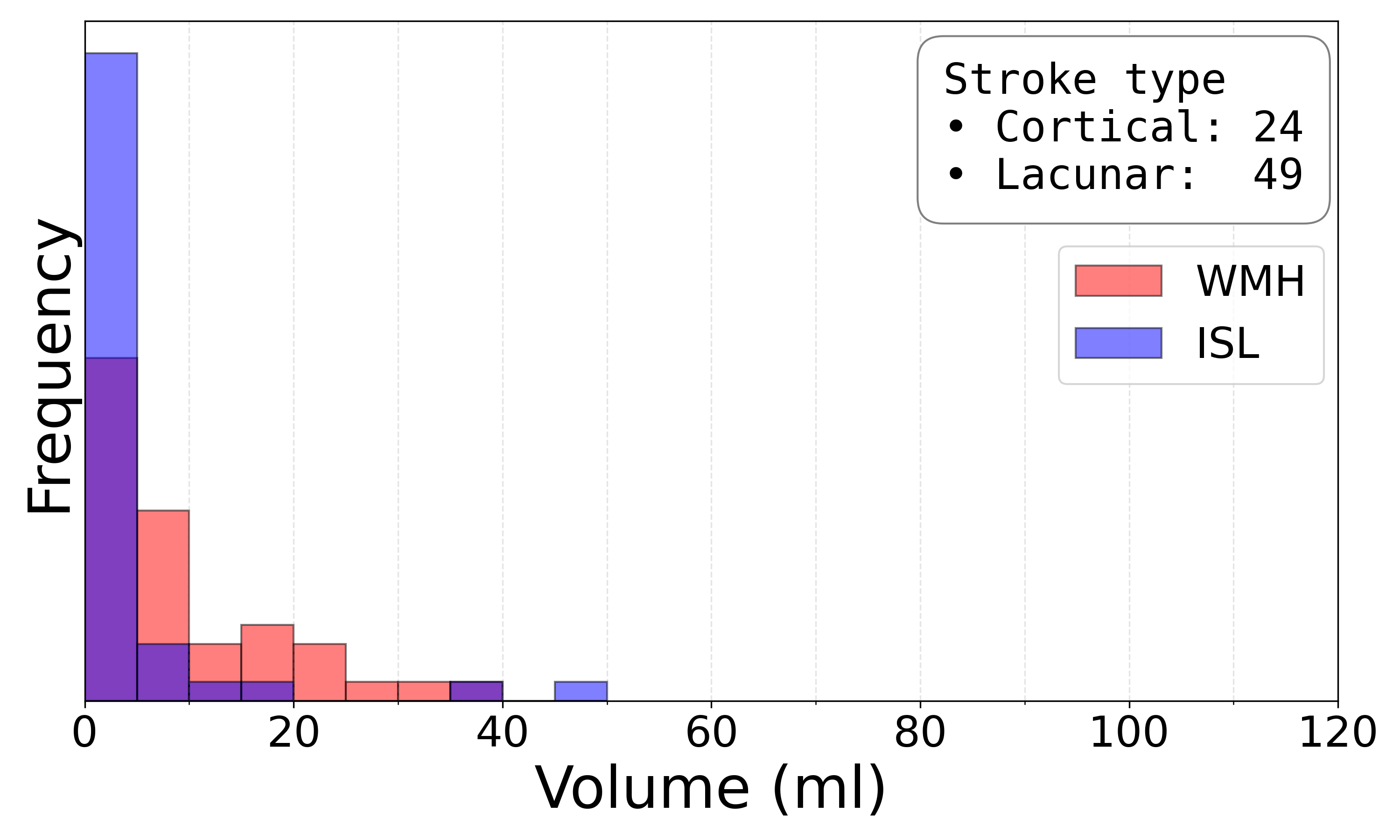}
     \end{subfigure}
     \hfill
     \begin{subfigure}[b]{0.32\textwidth}
         \centering
         \includegraphics[width=\textwidth]{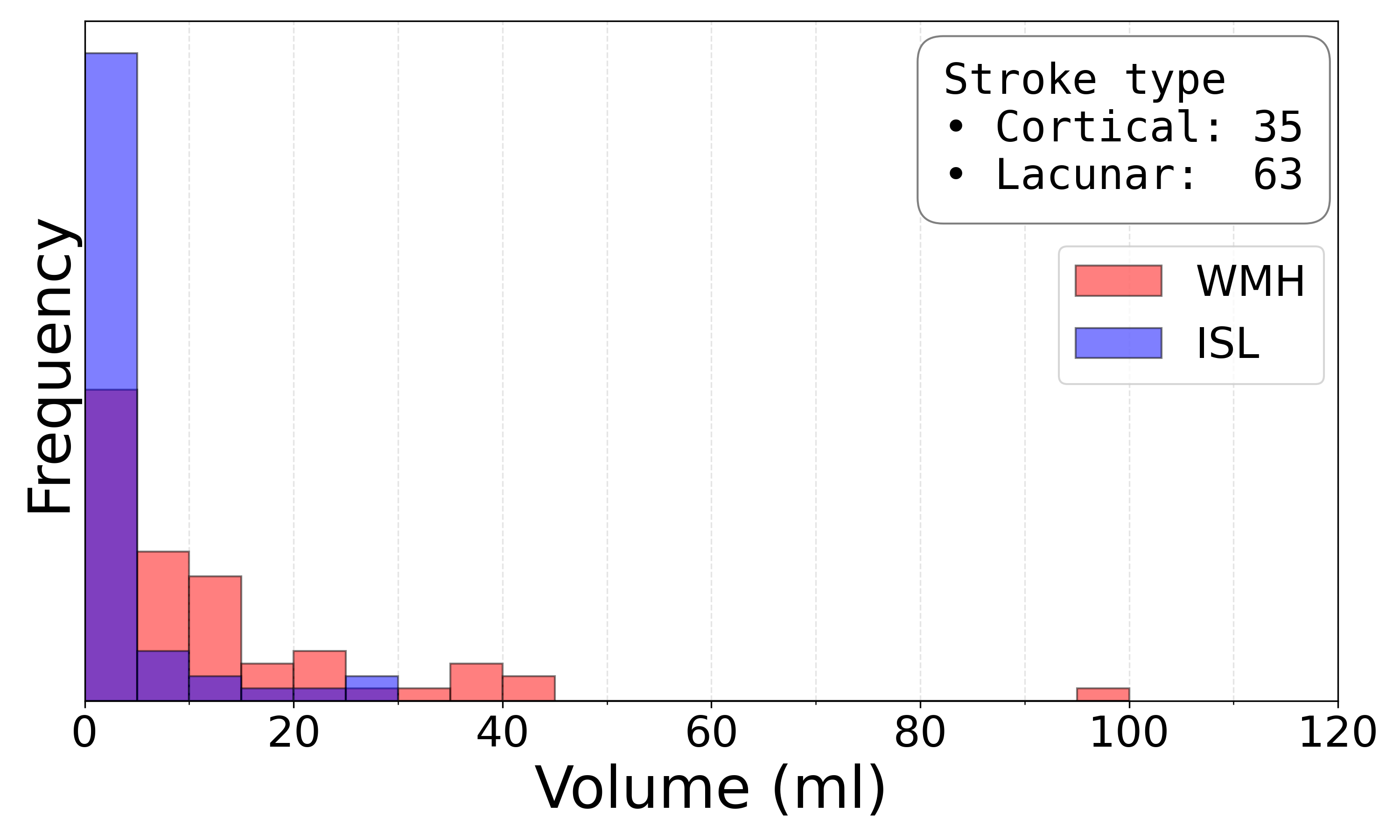}
     \end{subfigure}
        \caption{\textbf{Lesion volume distributions}. White matter hyperintensity (WMH) and ischaemic stroke lesion (ISL) volumes across the training (left), validation (middle) and test (right) set. The number of cortical or lacunar ISL are reported over all individual lesions since each subject may have more than one (or zero) ISL.}
        \label{ch5_fig:data_dist}
\end{figure}

Prior to model training, all scans underwent skull-stripping using SynthStrip \citep{hoopes2022synthstrip}, N4 bias-field correction \citep{tustison2010n4itk}, and resampling to a 1 mm isotropic resolution. Data was randomly partitioned into 103 training, 41 validation, and 62 test subjects (lesion distributions shown in \textbf{Figure \ref{ch5_fig:data_dist}}). 

\subsection{Hetero-modal training splits}
 
\begin{figure}[htbp]
    \centering
    \includegraphics[width=1\linewidth]{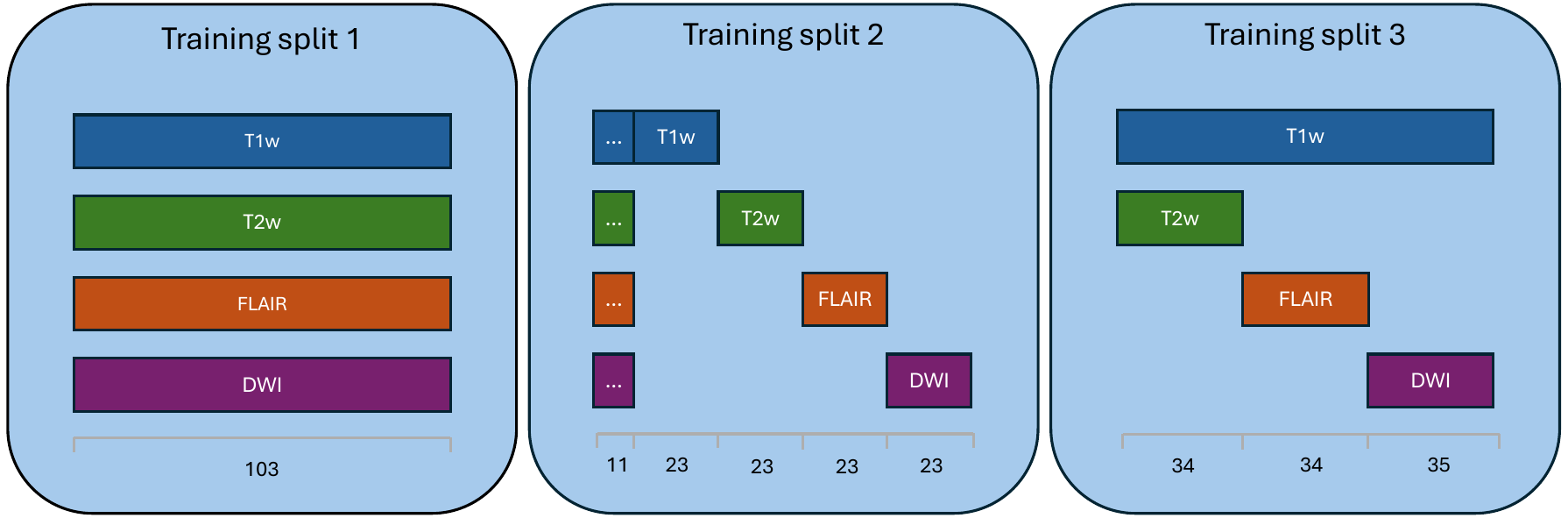}
    \caption{\textbf{Hetero-model training data splits}. Training data (n=103) distributions for the three different experimental setups (subject count on x-axis). Training split 1 has all four modalities for all subjects. Training split 2 has all four modalities for only 11 subjects, with the remaining subjects each having only a single modality (23 with T1w, 23 with T2w, 23 with FLAIR, and 23 with DWI). Training split 3 has T1w scans for all 103 subjects and each subject also has one addition sequence (34 with T2w, 34 with FLAIR, and 35 with DWI).}
    \label{ch5_fig:data}
\end{figure}

Clinical reality dictates that datasets frequently contain missing modalities. To expand the pool of usable training data for WMH and ISL segmentation, our goal is to leverage all available patient data to improve feature representation rather than discarding subjects with incomplete scans. To systematically evaluate this hetero-modal learning setting, we derived three different training data distributions from our dataset, illustrated in \textbf{Figure \ref{ch5_fig:data}}.

\subsubsection{Training split 1}
In this setup, all sequences are present for all training subjects. We used this modality-complete training split to evaluate and benchmark the overall effectiveness of each hetero-modal method against the modality specialist models (i.e., models that are trained and evaluated with a fixed combination of modalities). This split also served as the training distribution for our corruption augmentation-based experiments (described in section \ref{ch5_sec:corruption}), where we applied extreme data augmentation to enable the models to become robust not only to missing modalities, but to modalities corrupted with extreme levels of noise too.

\subsubsection{Training split 2}
In this highly uni-modal setup, a complete set of sequences is available for only a small subset (10\%) of the data, while the remaining subjects contribute only a single sequence. We designed this split to test whether hetero-modal learning can improve predictive performance on specific target sequences by leveraging additional, disjointed training data that a standard specialist model would otherwise be unable to access.

\subsubsection{Training split 3}
In this setup, we designated T1w as the ``anchor sequence'' due to its high prevalence in standard MRI protocols. All other sequences overlap with this T1w anchor, but never with each other. We used this distribution to investigate whether hetero-modal learning can train models capable of performing inference on novel, unseen combinations of sequences (e.g., FLAIR and DWI).

\subsection{Checkpoint selection} 
Because our primary objective is to evaluate the benefits of hetero-modal \textit{learning} (i.e., maximising training data) rather than hetero-modal inference (i.e., using a single set of weights for all possible inputs), we evaluated the model using the checkpoint that achieved the highest validation Dice Similarity Coefficient (DSC) for each specific inference modality setup. In other words, for each of the hetero-modal learning methods, we store 15 different model checkpoints from a single training run. This ensures the fairest possible comparison against the modality specialists, which naturally use their own optimal DSC checkpoints.

However, for the corruption augmentation experiments, we assume that corrupted modalities cannot be identified \textit{a priori} (otherwise, they would simply be treated as missing). Consequently, for those specific tests, we used the final model checkpoints---representing the state where the models have converged to a stable local minimum as the learning rate decays to zero.

\section{Methods} \label{ch5_sec:methods}

We implemented two established baseline methods for hetero-modal segmentation: a multi-channel U-Net \citep{ronneberger2015u} and U-HeMIS \citep{havaei2016hemis}. We also developed and evaluated our own architecture, the Multimodal Attention Router (MMAR), designed to enhance model robustness against corrupted modalities. We compared these hetero-modal methods to modality specialist baselines. This section details each of these architectures and our training methodologies. Code is available at: \url{https://github.com/Jesse-Phitidis/MultiModal}.

\subsection{Modality specialist baselines}
We trained modality-specialist baselines to serve as comparative benchmarks for the hetero-modal approaches. Each specialist model is trained and evaluated on a fixed, unique, modality combination. Consequently, these models cannot leverage the full multi-modal training data present in splits 2 and 3 (with the exception of the T1w specialist in training split 3). To prevent overfitting on their simpler learning tasks, the specialist baselines were configured with a standard 16, 32, 64, 128, 256, and 320 channels per resolution level, totalling approximately 16.7M parameters. 

\subsection{Multi-channel U-Net}
A multi-channel U-Net serves as the most straightforward hetero-modal method. Missing modalities are handled via zero-imputation (passed as channels of zeros), requiring the network to handle extreme distribution shifts within its very first layers. To ensure a fair comparison and provide the model with sufficient capacity to learn robustness to missing channels, we matched the parameter counts of the other hetero-modal methods (91M parameters) by configuring a large U-Net with 32, 64, 128, 256, 512, and 1024 channels at each respective resolution level.

\subsection{U-HeMIS}
As introduced in Section \ref{ch5_sec:intro}, HeMIS utilises an independent convolutional encoder for each modality. U-HeMIS extends this logic to the skip connections at every level of a U-Net architecture. At each resolution level, the network calculates the mean and variance across the feature maps extracted from the available modalities. These statistical representations are then concatenated along the channel dimension and processed by a single, shared convolutional decoder. For our experiments, we configured each independent encoder with 16, 32, 64, 128, 256, and 512 channels.

\subsection{Multimodal Attention Router (MMAR)}

\begin{figure}[htbp]
    \centering
    \includegraphics[width=1\linewidth]{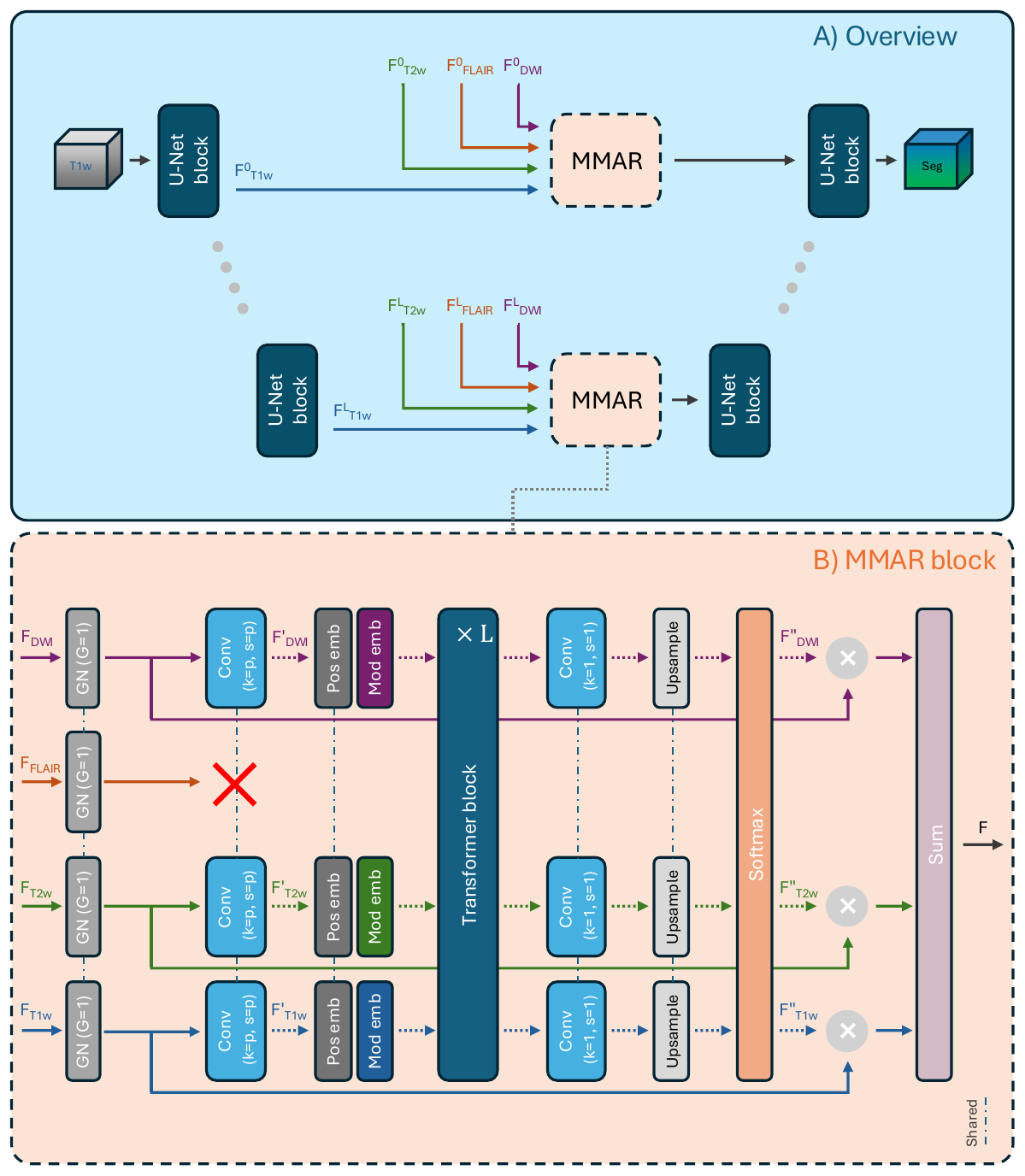}
    \caption{\textbf{Multimodal Attention Router (MMAR)-based architecture}. \textbf{A)} Overview of the multi-encoder, shared-decoder model (showing highest ($F^0$) and lowest ($F^L$) resolution levels for one encoder). MMAR blocks fuse features from each modality encoder prior to decoding. \textbf{B)} Detailed MMAR block illustrating fusion with one missing and three available modalities. Available features are normalised then patchified via a shared convolution and injected with positional and modality embeddings. Following processing by windowed self-attention transformer blocks, features are reshaped to their original dimensions. Finally, a softmax operation across the modality dimension generates spatial- and channel-wise attention maps, which are used to fuse the features via element-wise multiplication.}
    \label{ch5_fig:model}
\end{figure}

Motivated by the challenge presented by corrupted modalities which are not marked as missing, we propose the Multimodal Attention Router (MMAR) block. Our method, shown in \textbf{Figure \ref{ch5_fig:model}}, is inspired by U-HeMIS and SFusion \citep{liu2023sfusion}.

\subsubsection{Overview}
Our MMAR-based model replaces the statistical feature aggregation of U-HeMIS with a learnable dynamic feature router over the spatial and channel dimensions. The output is a single fused feature representation, with the important channels and spatial locations of each modality amplified, while irrelevant features are suppressed. This allows the MMAR to learn to down-weight features generated from corrupted modalities.

\subsubsection{MMAR block}

The MMAR block takes as input the encoded features from the $N$ modality encoders, ignoring the $M$ missing modalities. We denote the available, already encoded, input features as $\{F_i\}_{i=1}^{N-M}$, where each feature has $C$ channels and spatial dimensions $H \times W \times D$ (i.e., $F_i \in \mathbb{R}^{C \times H \times W \times D}$). The processing pipeline consists of three main stages:

\paragraph{Input normalisation and patchification}
To enhance the interpretability of the MMAR's attention weights and stabilise gradient magnitudes across the independent modality encoders, we applied group normalisation to the input features. This layer was specifically configured with a single group encompassing all channels and without any learnable affine parameters. Then, a shared 3D convolutional layer---with a kernel size and stride of $P \times P \times P$---patchifies and projects each input feature $F_i$ into an embedding dimension $E$. This operation yields the patch-level features $F'_i \in \mathbb{R}^{E \times \frac{H}{P} \times \frac{W}{P} \times \frac{D}{P}}$. 

\paragraph{Embedding and transformer processing}
Next, we flatten the spatial dimensions of each $F'_i$ to create a 1D sequence of tokens, each of length $\frac{H}{P} \cdot \frac{W}{P} \cdot \frac{D}{P}$. To retain spatial and modality-specific context, we add two learnable embeddings to these tokens: a positional embedding (shared across all modalities) and a modality embedding (shared across all tokens belonging to the same modality). The augmented tokens from all available modalities are then concatenated into a single sequence and passed through a series of $L$ transformer blocks. Each block consists of multi-head self-attention (MHSA) and a multi-layer perceptron (MLP). For computational efficiency, we employ windowed self-attention, which restricts each token to only attend to tokens within its local spatial window across all available modalities.

\paragraph{Reconstruction and fusion}
Following the transformer blocks, the processed sequence is reshaped back to its patchified spatial dimensions ($E \times \frac{H}{P} \times \frac{W}{P} \times \frac{D}{P}$). A $1 \times 1 \times 1$ convolution is then used to project the embedding dimension back to the original $C$ channels. We apply trilinear upsampling to these features to restore the original input resolution, producing the processed, full-resolution saliency features for each modality. We then apply a softmax function across the modality dimension to generate attention weight maps, denoted as $F''_i \in \mathbb{R}^{C \times H \times W \times D}$. The final fused feature representation, $F$, is computed as the element-wise weighted sum of the original input features with these weight maps:

\begin{equation}
    F = \sum^{N-M}_{i=1} F''_i \cdot F_i.
\end{equation}

\paragraph{Configuration details}
We use patch size $P=5$ at all resolution levels and a window size of $\text{min}(4, \text{number of patches at resolution})$. We set the embedding dimension equal to the number of channels at each resolution level (i.e., $E=C$) and the MLP hidden dimension to $4E$. We used $L=8$ transformer blocks, each with 8 attention heads. The modality-specific encoders mirrored the specialist baselines, utilising 16, 32, 64, 128, 256, and 320 channels per level.

\subsection{Implementation and training details}

\subsubsection{3D U-Net architecture}
All methods shared a common 3D U-Net backbone based on nnU-Net \citep{isensee2021nnu}, using two Conv-Norm-Act blocks per resolution level across six levels. We used $3\times3\times3$ convolutions, instance normalisation with affine scaling, and Leaky ReLU activations with a slope of 0.01. The encoder blocks incorporated residual connections, using strided convolutions for downsampling. Decoder upsampling was achieved via $2\times2\times2$ transposed convolutions (stride 2).

\subsubsection{Modality dropout}
To train the hetero-modal methods, we applied random modality dropout. For each training sample with $T$ available modalities (where $T$ may already be less than the total unique dataset modalities, $N$), we sampled an integer $D$ from the range [0, $T-1$]. We randomly dropped $D$ modalities, bringing the total number of missing modalities to $M=(N-T)+D$.

\subsubsection{Standard data augmentation}
We applied standard nnU-Net-style spatial and intensity augmentations: random affine transforms (prob=0.2, isotropic scaling in [0.7, 1.4], rotation in [-90$^\circ$, 90$^\circ$]); Gaussian noise (prob=0.15, std in [0, 0.316]); Gaussian blur (prob=0.1, sigma in [0.5, 1.5]); contrast adjustment (prob=0.15, multiplier in [0.65, 1.5]); gamma augmentation (prob=0.15, gamma in [0.7, 1.5]); low-resolution simulation (prob=0.25 per modality, factor in [1, 2]); and random axis flipping (prob=0.5). We also simulated motion, ghosting, spike, and bias field MRI artefacts (prob=0.05 each, via TorchIO \citep{perez2021torchio}). Post-augmentation, images were rescaled to [0, 1], saturating at the 0.5 and 99.5 percentiles.

\subsubsection{Corruption augmentation} \label{ch5_sec:corruption}
\begin{figure}
    \centering
    \includegraphics[width=0.75\linewidth]{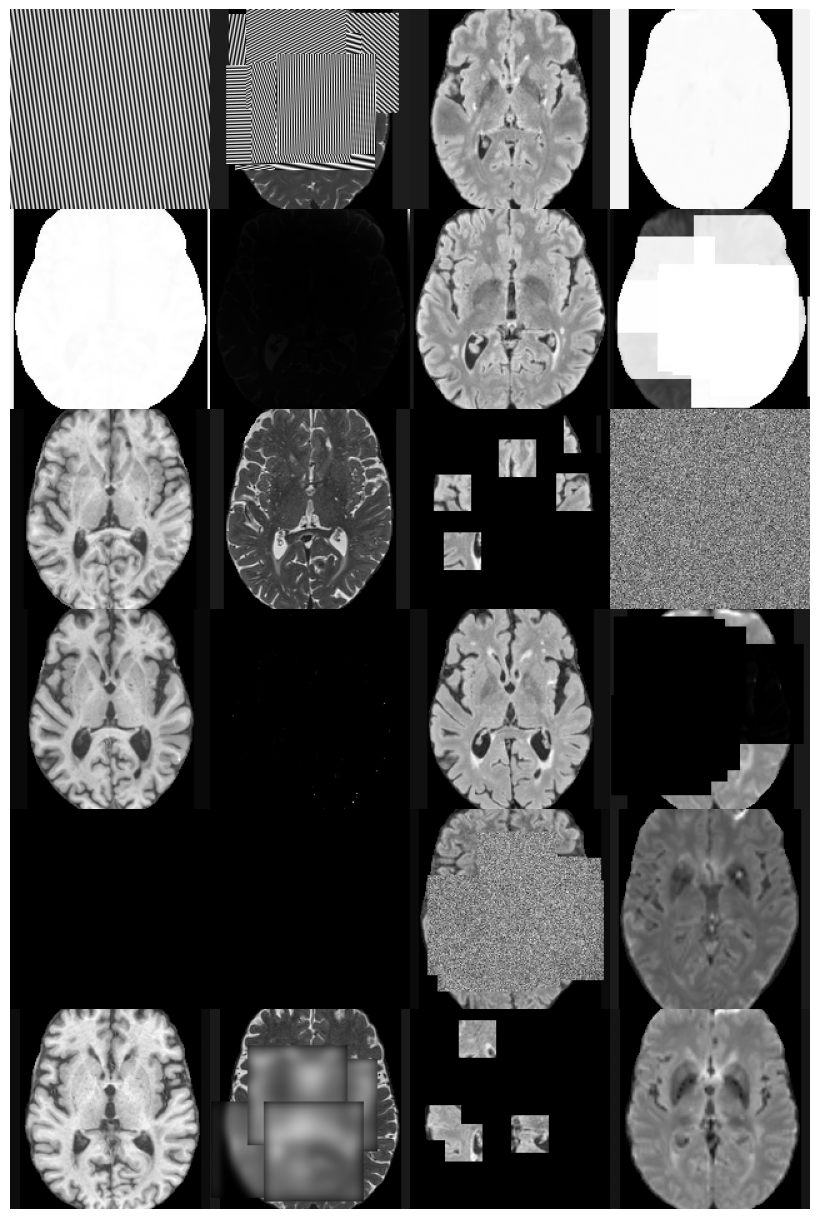}
    \caption{\textbf{Dropout and corruption augmentation}. Data augmentation applied to batches of T1w, T2w, FLAIR, and DWI (left to right) MRI scans during training (for the corruption-based experiments).}
    \label{ch5_fig:train_corruption}
\end{figure}

To equip the hetero-modal models with resilience to unidentified corrupted modalities, we performed some of our experiments with an additional layer of extreme  data augmentation severe enough to render a scan almost or completely unusable. During training for these experiments, we reassigned  (with 50\% probability per scan) some of the $D$ scans that were initially selected to undergo modality dropout, to undergo corruption instead. For each corrupted scan, the corruption transform was randomly selected as one of the following: gaussian noise (std=3); bias field (polynomial degree 3, coefficients in [0.4, 0.9]); low contrast (multiplier in [0, 0.1]); high contrast (multiplier in [10, 30]); Gaussian blur (sigma in [2, 6] per axis); Gibbs noise (alpha in [0.75, 1]); K-Space spike (intensity in [15, 18]); Rician noise (std=1). We randomly performed each of these corruptions on the entire scan, or on random patches of the scan (3-10 patches, each of size 40-80 voxels in each axis). We implemented these transformations using the MONAI library \citep{cardoso2022monai}. \textbf{Figure \ref{ch5_fig:train_corruption}} shows examples.

\subsubsection{Loss functions, optimisation, and schedule}
Models were optimised using a combination of Dice and Cross-Entropy loss:

\begin{equation}
    \mathcal{L}_{seg} = 0.5\mathcal{L}_{DSC} + 0.5\mathcal{L}_{CE}.
\end{equation}

We applied deep supervision across the four highest-resolution decoder output levels ($l$), scaling the loss with exponentially decreasing weights:

\begin{equation}
    w_l = \frac{2^{-l}}{\sum^3_{i=0} 2^{-i}}.
\end{equation}

We trained all models for 2000 epochs (defining an epoch as 103 subjects, which is the total number of training subjects available when using all modalities) using the AdamW optimiser (base learning rate 0.0003, weight decay 0.01) and a batch size of 2. The learning rate followed a linear warm-up (20 epochs) into a cosine decay schedule. Validation checkpoints were evaluated every 50 epochs. For training split 2, models trained exclusively on the full-modality subset for the initial 100 epochs to stabilise early learning.

\subsubsection{MMAR-specific training details}
When training the MMAR-based model with corruption augmentation, we introduced an auxiliary loss term to explicitly penalise the router for assigning high attention weights to corrupted modalities. For each MMAR block level $l$, the auxiliary loss over the set of corrupted modalities $\mathcal{K}$ is:

\begin{equation}
\mathcal{L}^{(l)}_{aux} = \frac{1}{|\mathcal{K}|}\sum_{i \in \mathcal{K}} \mathcal{L}_{CE}(\operatorname{mean}(F''_i), 0)
\end{equation}

The total loss $\mathcal{L}_{MMAR}$ for the network (with $L=6$ encoder levels and $\alpha=0.9$) is defined as:

\begin{equation}
    \mathcal{L}_{aux} = \frac{1}{L} \sum_{l=1}^L \mathcal{L}^{(l)}_{aux}
\end{equation}

\begin{equation}
    \mathcal{L}_{MMAR} = \frac{N-M}{N} \cdot (\alpha \mathcal{L}_{seg} + (1 - \alpha) \mathcal{L}_{aux}).
\end{equation}

The inclusion of this auxiliary loss relies on the assumption that none of the training scans are actually corrupted to begin with.

Additionally, we observed that scaling the total loss by the fraction of available modalities per sample (e.g., scaling by 0.5 if only T1w and FLAIR were present) improves MMAR training stability on splits 2 and 3. We hypothesise that this reduces gradient interference between  uni/few-modal samples caused by the high variance in feature distributions produced by the dynamic routing mechanism, when trained on our small dataset. This scaling negatively impacts the validation metrics for the multi-channel U-Net and U-HeMIS, so it was applied only to the MMAR.

\subsection{Evaluation}

\subsubsection{Metric}
We evaluated segmentation performance using the Dice Similarity Coefficient (DSC), defined as:

\begin{equation}
    DSC = \frac{2|Y \cap \hat{Y}|}{|Y| + |\hat{Y}|}
\end{equation}

where $Y$ and $\hat{Y}$ represent the ground truth and predicted segmentation masks, respectively.

\subsubsection{Test time corruption}

\begin{figure}[htbp]
    \centering
    \includegraphics[width=1\linewidth]{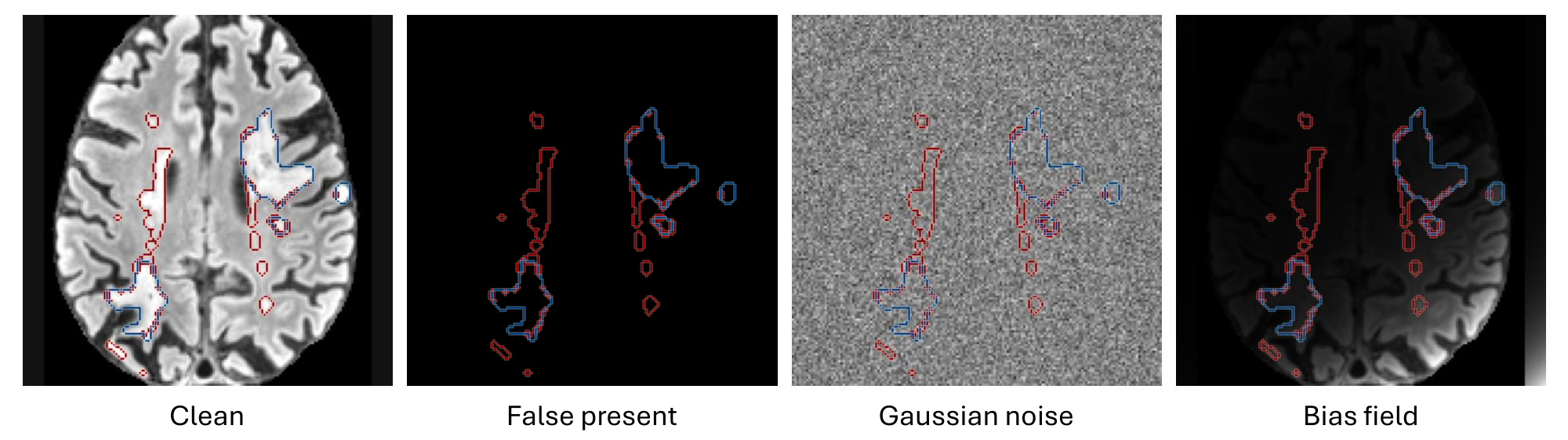}
    \caption{\textbf{Test time corruption examples}. Transformations applied on an example FLAIR scan from the validation set. The image with bias field corruption is too dark to see by eye, so it has been rescaled for visualisation purposes only. White matter hyperintensities (WMH) in red and ischaemic stroke lesions (ISL) in blue.}
    \label{ch5_fig:corruption_example}
\end{figure}

Because our dataset lacks truly corrupted images, we simulated catastrophic degradation during testing using Gaussian noise (std=10) and bias field artefacts (polynomial degree 2, coefficients in [0.7, 0.9]). These parameters differ substantially from the training augmentations to reduce evaluation bias. We also introduced a ``false present'' scenario---setting an entire scan to zero except for a single corner voxel---simulating a missing scan that bypasses automated missing-data detection. \textbf{Figure \ref{ch5_fig:corruption_example}} shows an example of each test-time corruption type.

\section{Results and discussion}

\subsection{Modality specialists win when trained on fully multi-modal data}

\textbf{Table \ref{ch5_tab:tab_split_1}} demonstrates that in an ideal training scenario---where all subjects possess all modalities---the specialist U-Nets achieve the strongest mean macro-average performance, showing a significant improvement over the multi-channel U-Net. Although U-HeMIS or MMAR yielded the highest Dice Similarity Coefficient (DSC) in 6 out of 15 inference modality combinations, these improvements over the specialist U-Net were not statistically significant.

\begin{table}[htbp]
    \centering
    \caption{\textbf{Hetero-modal learning experiments --- training split 1}. Ticks represent present and crosses represent missing modalities. Superscripts $a$, $b$, $c$, and $d$ are shorthand for Spec., U-Net, U-HeMIS, and MMAR respectively and DSC$^x$ denotes a DSC which is significantly better ($p<0.05$) than the DSC achieved by method $x$ (one-sided Wilcoxon signed-rank test with Holm-Bonferroni correction). DSC: Dice similarity coefficient.}
    \includegraphics[width=1\linewidth]{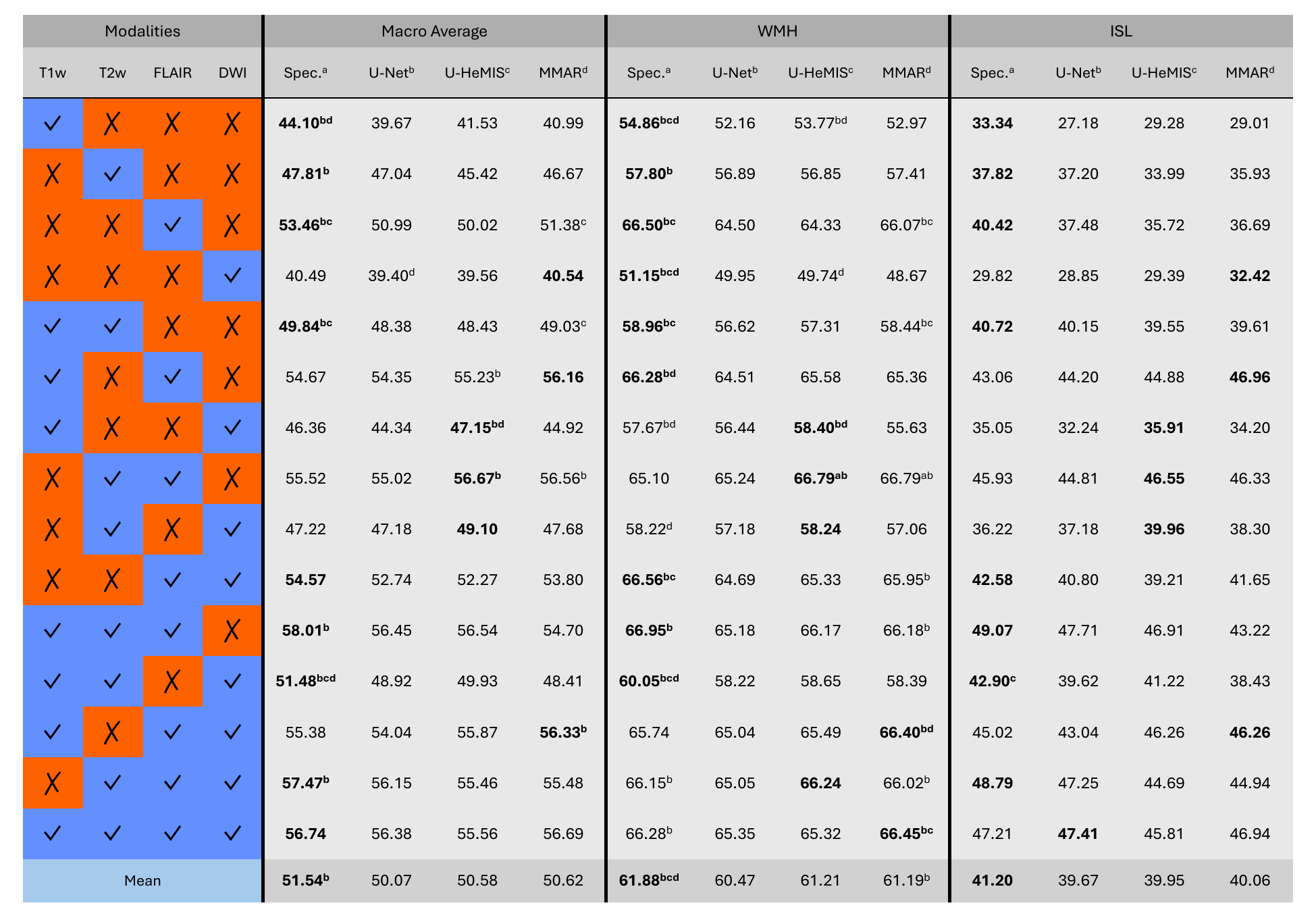}
    \label{ch5_tab:tab_split_1}
\end{table}

Evaluating specialist models trained on single modalities reveals the independent predictive value of each MRI sequence. As anticipated, FLAIR is the most informative modality for both WMH and ISL segmentation, while DWI is the least informative, and T2w outperforms T1w. The highest macro-average DSC on this dataset (58.01\%) was achieved by the specialist model using T1w, T2w, and FLAIR. Interestingly, the second-best result utilised T2w, FLAIR, and DWI. While one might expect the integration of all four modalities to yield peak performance, these results suggest that three complementary modalities are sufficient. Adding a fourth may actually destabilise model fitting on our small dataset and degrade inference robustness by introducing additional modes of distribution shift among the multi-modal features. For the hetero-modal methods, the multi-channel U-Net peaked at 56.45\% (using T1w, T2w, and FLAIR), whereas U-HeMIS and MMAR achieved their highest scores of 56.67\% (T2w and FLAIR inference) and 56.69\% (full modality inference), respectively.

\subsection{Hetero-modal models win when trained on highly uni-modal data}

\textbf{Table \ref{ch5_tab:tab_split_2}} highlights the value of hetero-modal learning for maximising training data utilisation when dealing with mostly non-overlapping MRI sequences. Across almost all modality setups, the hetero-modal approaches outperformed the specialist models, except MMAR on T1w and U-HeMIS on DWI. The mean macro-average result was significantly better for all hetero-modal models. Furthermore, in every scenario involving more than one inference modality, all hetero-modal approaches yielded statistically significant improvements over the specialist U-Nets.

\begin{table}[htbp]
    \centering
    \caption{\textbf{Hetero-modal learning experiments --- training split 2}. Ticks represent present and crosses represent missing. Superscripts $a$, $b$, $c$, and $d$ are shorthand for Spec., U-Net, U-HeMIS, and MMAR respectively and DSC$^x$ denotes a DSC which is significantly better ($p<0.05$) than the DSC achieved by method $x$ (one-sided Wilcoxon signed-rank test with Holm-Bonferroni correction). DSC: Dice similarity coefficient.}
    \includegraphics[width=1\linewidth]{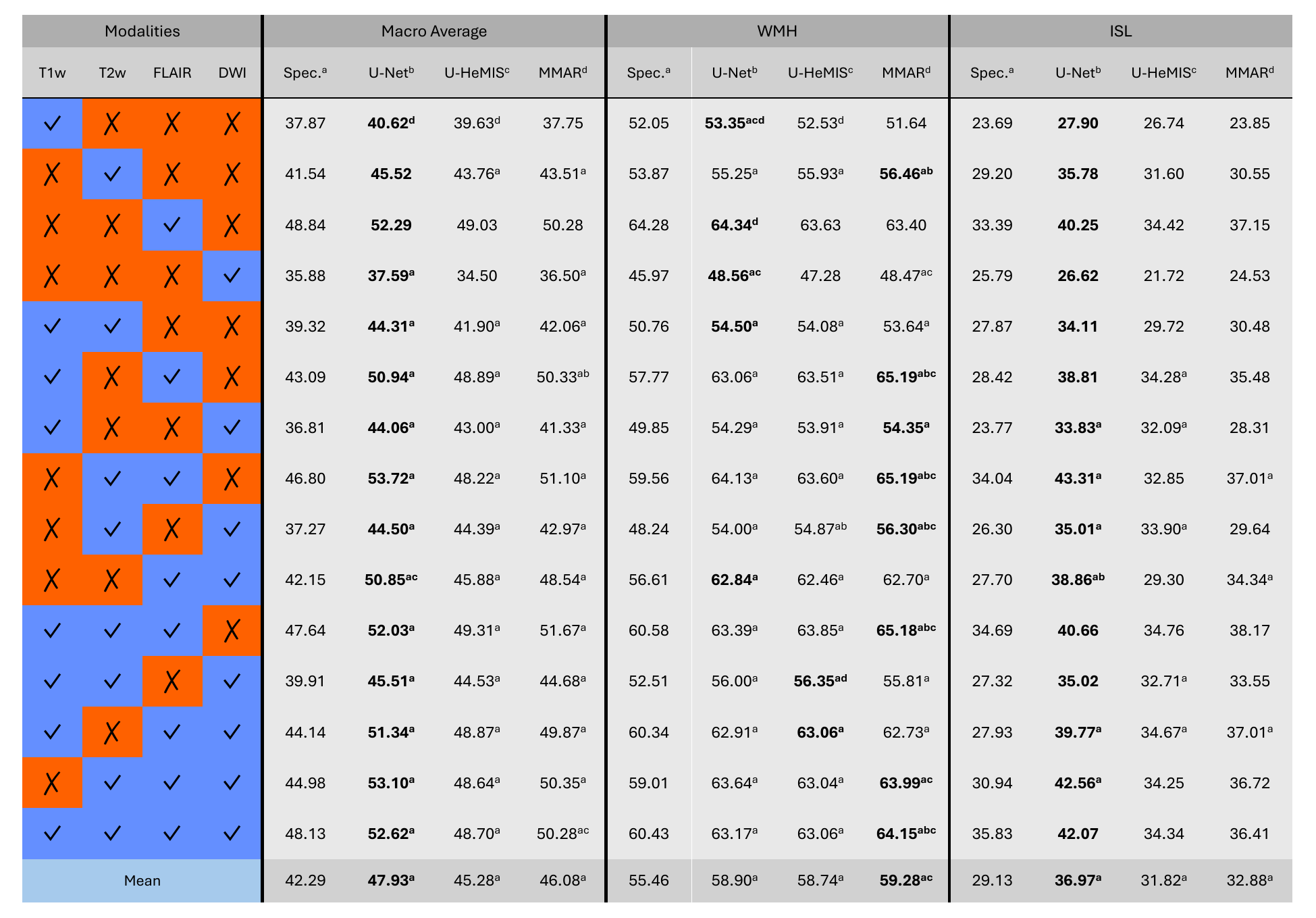}
    \label{ch5_tab:tab_split_2}
\end{table}

In contrast to the complete-data scenario of split 1, the multi-channel U-Net generally outperformed both U-HeMIS and MMAR in split 2, though the mean differences were not statistically significant. The multi-channel U-Net's advantage stems primarily from its strong ISL segmentation results, achieving the highest DSC across all inference combinations. Meanwhile, the MMAR method achieved the best WMH DSC, significantly outperforming U-HeMIS, but not the multi-channel U-Net.

We hypothesise that this dynamic occurs because the independent encoders of U-HeMIS and MMAR are exposed to less overall pathological variation, which is a critical factor for successful ISL segmentation. In this highly uni-modal data split, the benefit of sharing features across the entire multi-channel U-Net appears to outweigh any drawbacks of early fusion via zero-imputation. If this hypothesis is true, then further experimentation is required to explore the relative impact of the total volume of multi-modal data and the ratio of multi-modal to uni-modal data.

\subsection{Hetero-modal models perform well on unseen modality combinations}

\textbf{Table \ref{ch5_tab:tab_split_3}} outlines the results for training split 3, further reinforcing the benefits of hetero-modal learning. The specialist models only achieved the best result during T1w inference---the single ``anchor'' modality present across all training samples. Conversely, the MMAR architecture demonstrated the strongest performance on 6 out of 8 novel modality combinations. Its mean macro-average score was significantly higher than that of U-HeMIS, though not significantly better than the multi-channel U-Net. We attribute this success to the MMAR’s locally adaptive, dynamic feature weighting, which handles novel modality combinations more robustly than the fixed statistical merging used in U-HeMIS.

\begin{table}[htbp]
    \centering
    \caption{\textbf{Hetero-modal learning experiments --- training split 3}. Ticks represent present and crosses represent missing. Superscripts $a$, $b$, $c$, and $d$ are shorthand for Spec., U-Net, U-HeMIS, and MMAR respectively and DSC$^x$ denotes a DSC which is significantly better ($p<0.05$) than the DSC achieved by method $x$ (one-sided Wilcoxon signed-rank test with Holm-Bonferroni correction). DSC: Dice similarity coefficient.}
    \includegraphics[width=1\linewidth]{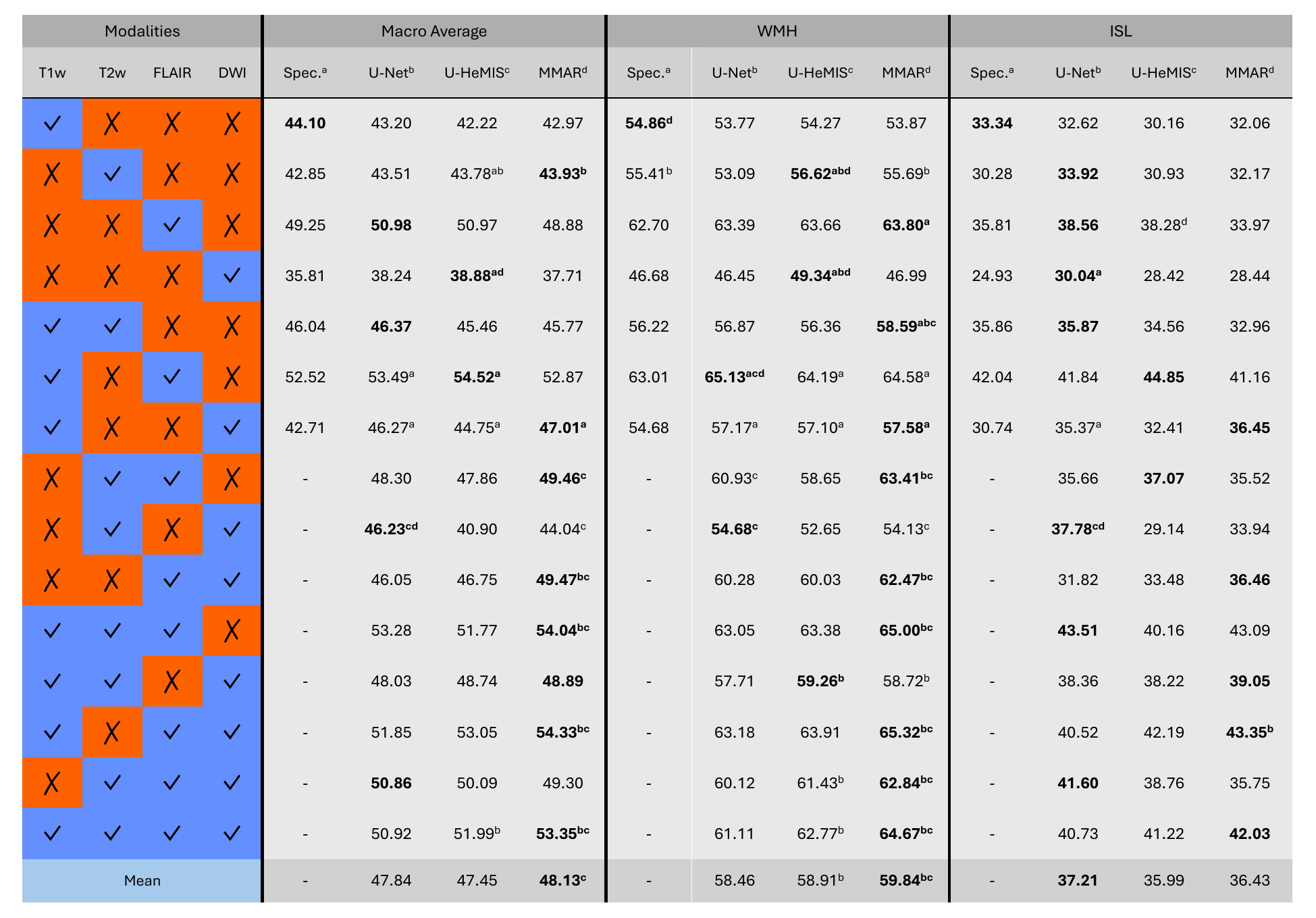}
    \label{ch5_tab:tab_split_3}
\end{table}

While MMAR is the most promising approach for inference on unseen modality combinations, the absolute highest macro-average DSC (54.52\%) on the test set was achieved by U-HeMIS using T1w and FLAIR. MMAR secured the second and third best overall results---54.33\% (T1w, FLAIR, DWI) and 54.04\% (T1w, T2w, FLAIR)---both of which were novel combinations.

When the crucial FLAIR sequence was unavailable, MMAR achieved the peak performance of 48.89\% using a novel combination of T1w, T2w, and DWI. This specific combination also yielded the best non-FLAIR results for the multi-channel U-Net and U-HeMIS. This underscores the power of hetero-modal learning for navigating missing modality scenarios at inference time: without it, the next best non-FLAIR result was only 47.01\%, achieved by MMAR using an overlapping combination (T1w and DWI).

\subsection{MMAR outperforms established hetero-modal models on corrupted data}

The final experiments evaluated model robustness against severe data degradation. We applied corruption augmentation during training on split 1 and tested under four scenarios: completely missing modalities, ``false present'' modalities (missing but unidentified as such), catastrophic Gaussian noise, and severe bias fields.

As shown in \textbf{Table \ref{ch5_tab:tab_split_1_c_missing}}, when predicting with explicitly missing modalities, the multi-channel U-Net and MMAR performed slightly worse than they did without corruption augmentation, whereas U-HeMIS showed slight improvement.

\begin{table}[htbp]
    \centering
    \caption{\textbf{Corruption experiments --- missing modalities}. Training split 1 trained with corruption augmentation. Ticks represent present and clean modalities and crosses represent missing modalities. Specialist models (Spec.) not trained with corruption augmentation or included in statistical comparison, but are included for reference. Superscripts $b$, $c$, and $d$ are shorthand for U-Net, U-HeMIS, and MMAR respectively and DSC$^x$ denotes a DSC which is significantly better ($p<0.05$) than the DSC achieved by method $x$ (one-sided Wilcoxon signed-rank test with Holm-Bonferroni correction). DSC: Dice similarity coefficient.}
    \includegraphics[width=1\linewidth]{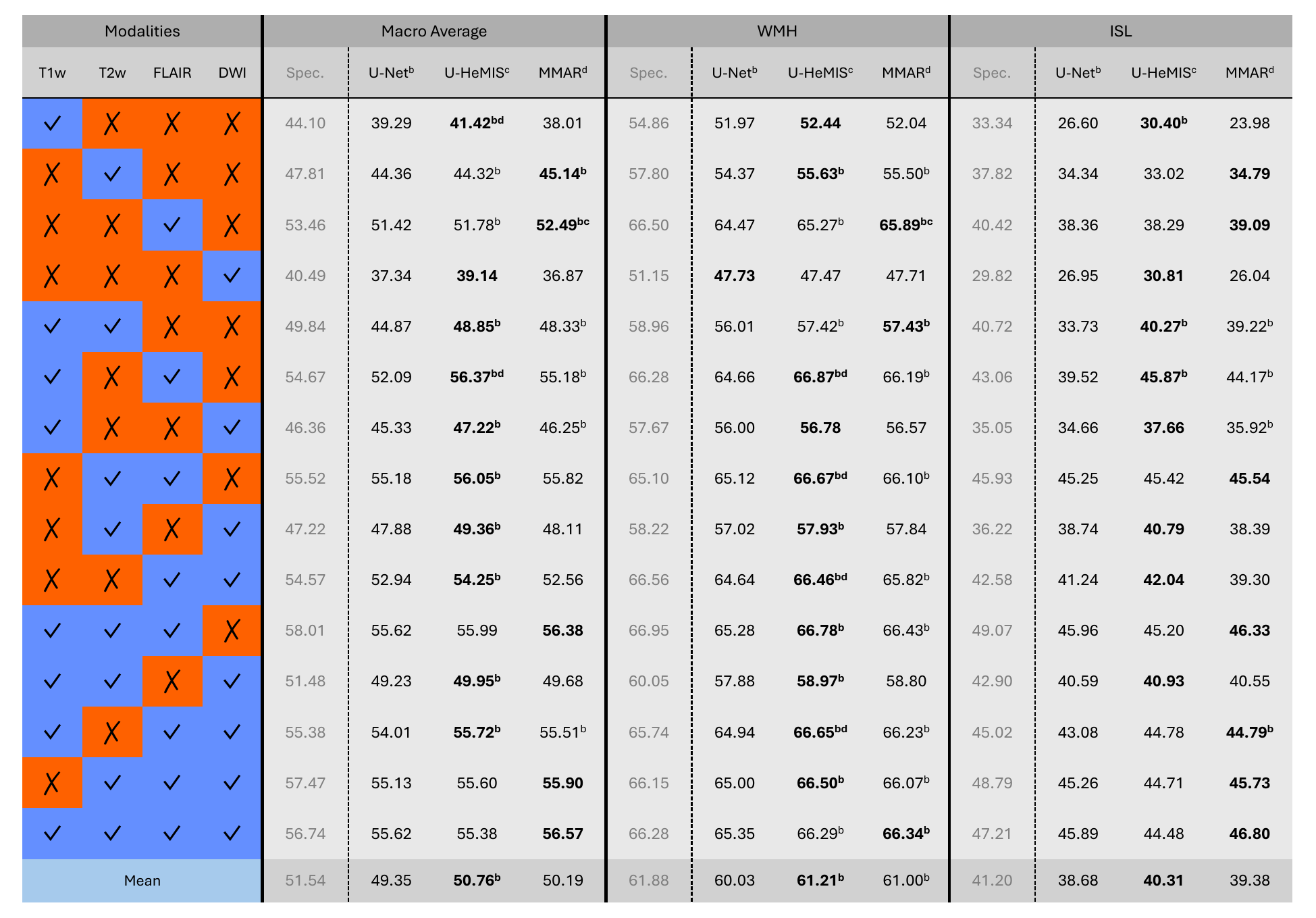}
    \label{ch5_tab:tab_split_1_c_missing}
\end{table}

However, the results shift under the remaining three corruption scenarios (\textbf{Tables \ref{ch5_tab:tab_split_1_c_false_present}, \ref{ch5_tab:tab_split_1_c_gaussian}, and \ref{ch5_tab:tab_split_1_c_bias_field}}). Faced with false presents, Gaussian noise, or severe bias fields, the MMAR consistently outperformed both the multi-channel U-Net and U-HeMIS for both WMH and ISL segmentation, and exhibited minimal performance degradation when transitioning from explicitly missing to corrupted modality testing.

\begin{table}[htbp]
    \centering
    \caption{\textbf{Corruption experiments --- false present modalities}. Training split 1 trained with corruption augmentation. Ticks represent present and clean modalities and crosses represent ``false present'' (single non-zero voxel) modalities. Specialist models (Spec.) not trained with corruption augmentation or included in statistical comparison, but are included for reference. Superscripts $b$, $c$, and $d$ are shorthand for U-Net, U-HeMIS, and MMAR respectively and DSC$^x$ denotes a DSC which is significantly better ($p<0.05$) than the DSC achieved by method $x$ (one-sided Wilcoxon signed-rank test with Holm-Bonferroni correction). DSC: Dice similarity coefficient.}
    \includegraphics[width=1\linewidth]{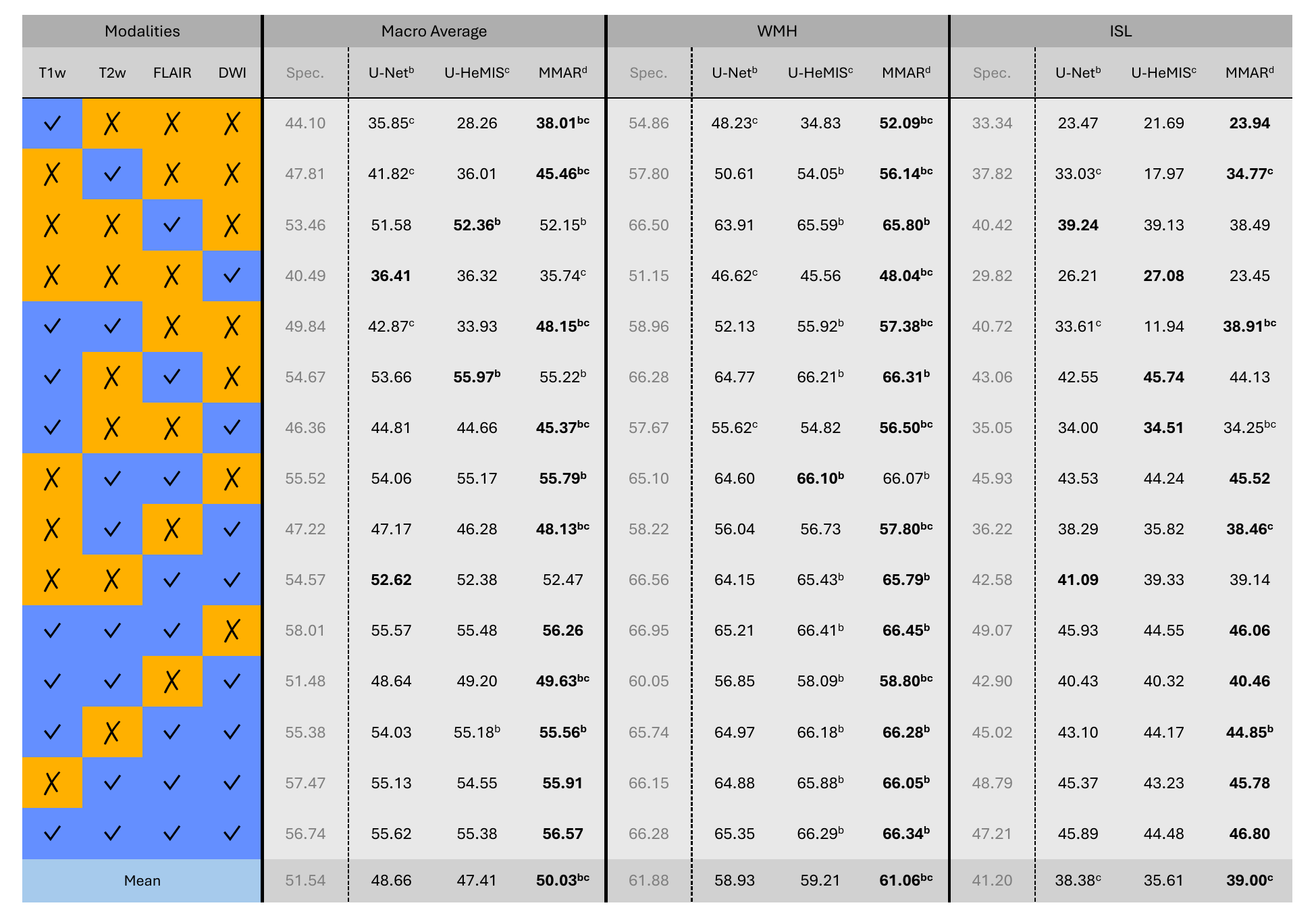}
    \label{ch5_tab:tab_split_1_c_false_present}
\end{table}

\begin{table}[htbp]
    \centering
    \caption{\textbf{Corruption experiments --- modalities corrupted by Gaussian noise}. Training split 1 trained with corruption augmentation. Ticks represent present and clean modalities and crosses represent modalities corrupted by catastrophic Gaussian noise. Specialist models (Spec.) not trained with corruption augmentation or included in statistical comparison, but are included for reference. Superscripts $b$, $c$, and $d$ are shorthand for U-Net, U-HeMIS, and MMAR respectively and DSC$^x$ denotes a DSC which is significantly better ($p<0.05$) than the DSC achieved by method $x$ (one-sided Wilcoxon signed-rank test with Holm-Bonferroni correction). DSC: Dice similarity coefficient.}
    \includegraphics[width=1\linewidth]{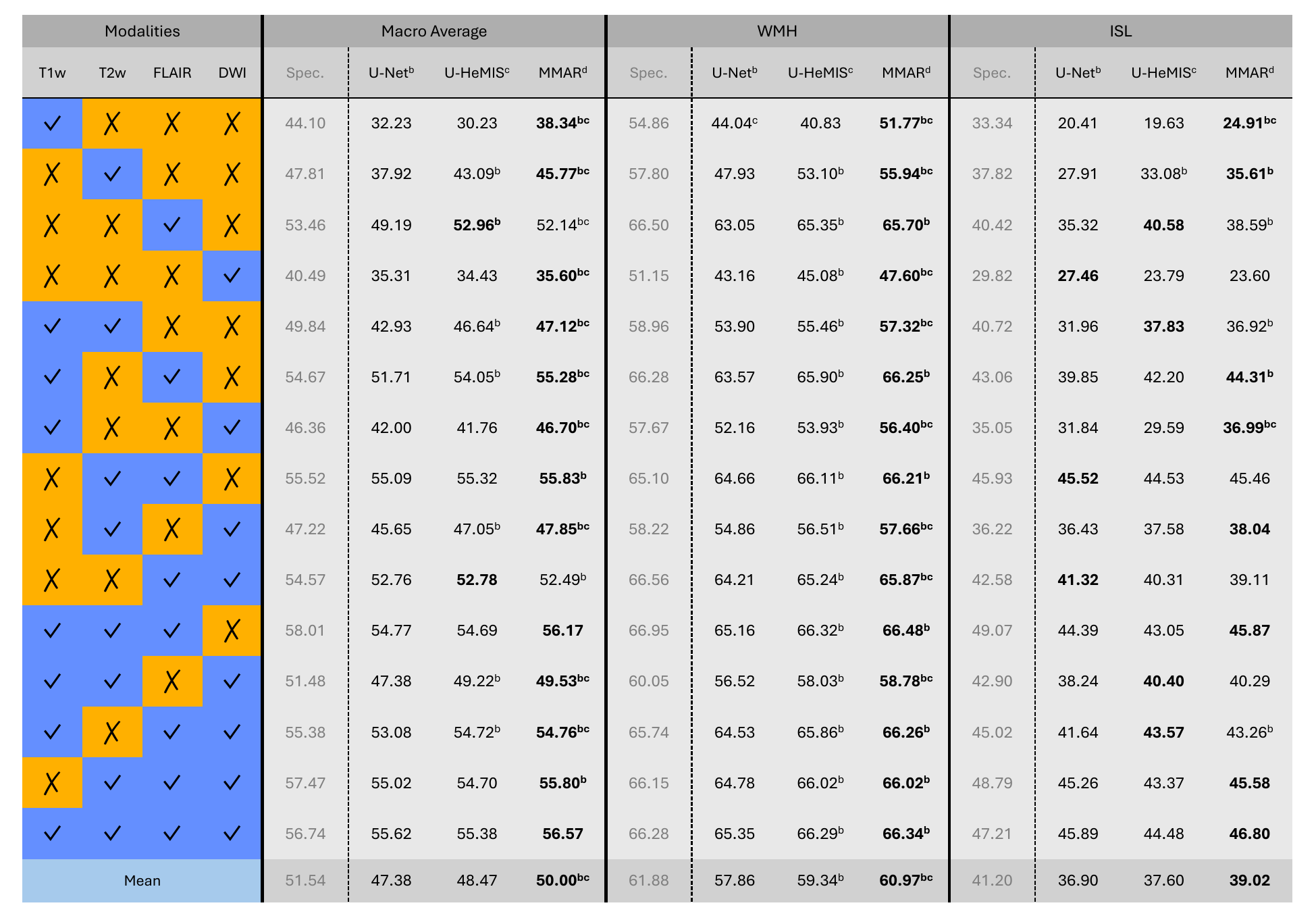}
    \label{ch5_tab:tab_split_1_c_gaussian}
\end{table}

\begin{table}[htbp]
    \centering
    \caption{\textbf{Corruption experiments --- modalities corrupted by bias field artefacts}. Training split 1 trained with corruption augmentation. Ticks represent present and clean modalities and crosses represent modalities corrupted by severe bias field artefacts. Specialist models (Spec.) not trained with corruption augmentation or included in statistical comparison, but are included for reference. Superscripts $b$, $c$, and $d$ are shorthand for U-Net, U-HeMIS, and MMAR respectively and DSC$^x$ denotes a DSC which is significantly better ($p<0.05$) than the DSC achieved by method $x$ (one-sided Wilcoxon signed-rank test with Holm-Bonferroni correction). DSC: Dice similarity coefficient.}
    \includegraphics[width=1\linewidth]{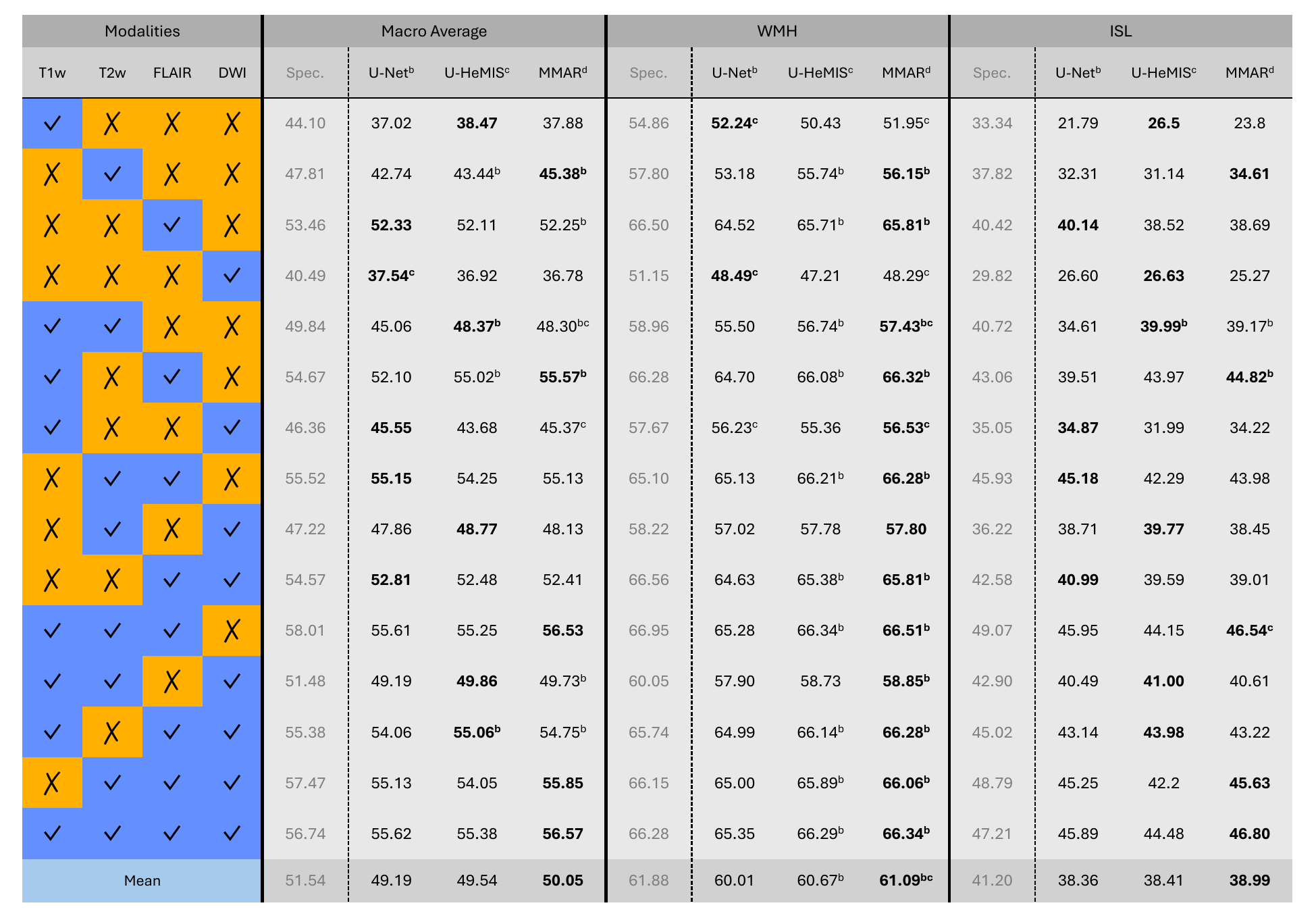}
    \label{ch5_tab:tab_split_1_c_bias_field}
\end{table}

We attribute the MMAR's success to its router-based architecture and auxiliary router loss. During training, we observed the router loss drop rapidly and remain low. This mechanism drastically reduces the gradient flow to any encoder processing a corrupted image, protecting that encoder from wasting learning capacity on becoming robust to catastrophic noise.

\subsection{MMAR: attention router analysis}

\begin{figure}[htbp]
    \centering
    \includegraphics[width=1\linewidth]{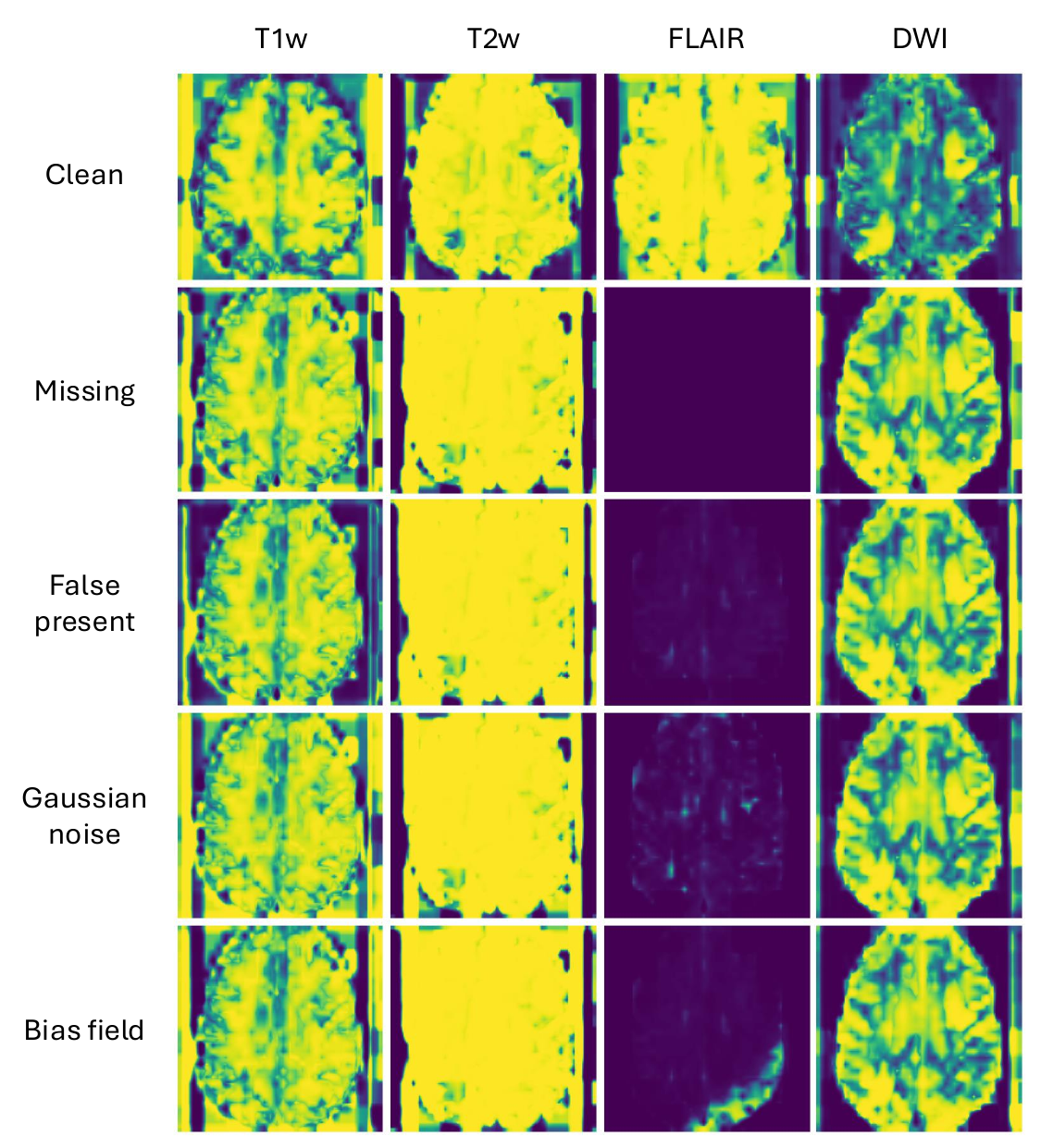}
    \caption{\textbf{Visualisation of MMAR attention weights}. Visualised via maximum intensity projection over the channels of the MMAR attention weights at the highest resolution level of the encoder when a FLAIR scan from the validation set is subjected to different missingness and corruption scenarios. The figure shows that global corruption (false present and Gaussian noise) results in similar attention maps to truly missing, and that local corruption (bias field) leads to only locally reduced attention.}
    \label{ch5_fig:mips}
\end{figure}

\textbf{Figure \ref{ch5_fig:mips}} visualises the maximum intensity projection (across all 16 channels) of the router's output weights at the highest encoder resolution level, demonstrating how the network responds when the FLAIR scan is subjected to various missingness and corruption scenarios.

The first row illustrates the baseline output when all modalities are clean and present. Here, the router is highly selective about where it assigns high attention weights to the DWI scan. The two prominent bright areas strongly align with the locations of the ISL, while weights remain comparatively suppressed outside these regions. This aligns with clinical intuition: DWI is a powerful modality for identifying acute ISL (often termed ``DWI-positive'') but generally lacks anatomical detail and is acquired at a lower resolution.

When FLAIR is completely missing, the attention distributions shift significantly---most notably for T2w and DWI. Features previously dominated by the FLAIR sequence in the fused representation are drawn from the remaining available modalities. Specifically, the network relies more heavily on the DWI encoder for grey matter features and the T2w encoder for background features.

When the FLAIR scan is subjected to catastrophic corruption (Gaussian noise, false present), the router assigns it negligible weight. The resulting attention maps for T1w, T2w, and DWI closely resemble those of the explicitly missing FLAIR scenario. This demonstrates that the MMAR naturally learns to treat severely corrupted modalities as if they were missing. Under bias field corruption, the router leverages its spatial adaptivity to selectively allow FLAIR features to pass into the hetero-modal fusion only in regions where the scan remains less affected by the artefact.

\begin{figure}
    \centering
    \includegraphics[width=1\linewidth]{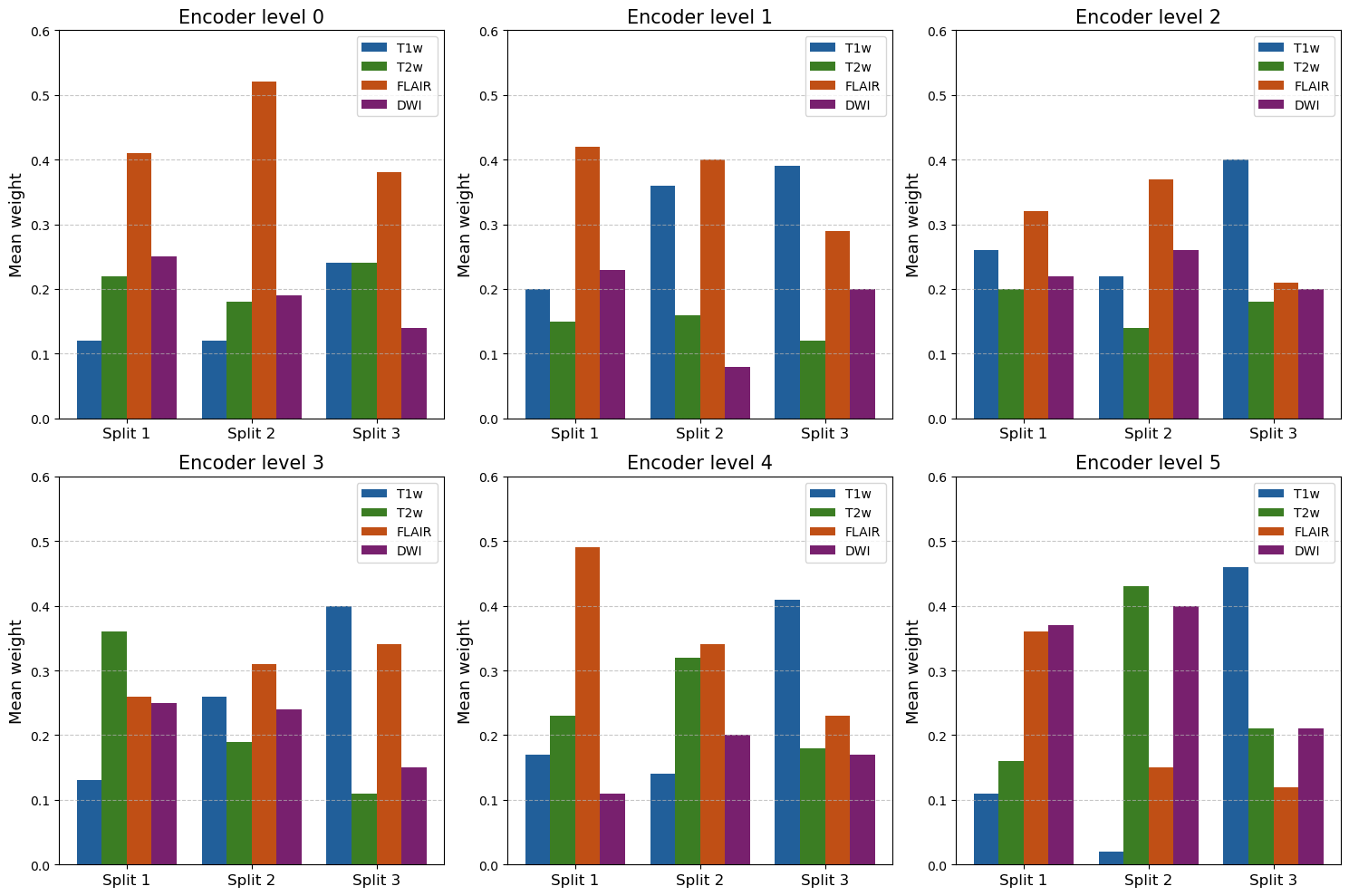}
    \caption{\textbf{Modality distribution of MMAR attention weights}. The figure shows the mean MMAR attention weights for each modality on the validation set when performing full modality inference, when trained on training data splits 1, 2, and 3. The final checkpoints were used for all models. Encoder level 0 is the highest resolution level and 5 is the bottleneck.}
    \label{ch5_fig:bar_chart}
\end{figure}

\textbf{Figure \ref{ch5_fig:bar_chart}} plots the mean router weights on the validation set at each encoder level during full-modality inference (using the final model checkpoint), comparing models trained across the three different data splits (without corruption augmentation). Across training splits 1 and 2, FLAIR dominates the three highest resolution levels (levels 0-2).

However, these plots reveal that the underlying training data distribution influences the weighting strategy learned by the router. For instance, in training split 3---where T1w scans are the ``anchor'' modality and far more abundant than the others---the router learns to prioritise T1w features. Here, FLAIR only retains its dominance at the absolute highest resolution level. This result is a logical consequence of the network maximising the utility of the T1w encoder, which naturally becomes a stronger feature extractor due to the higher T1w representation in the training data.

\section{Conclusion}

This study establishes the first benchmark for the joint hetero-modal segmentation of co-occurring white matter hyperintensities (WMH) and ischaemic stroke lesions (ISL). By leveraging a unique, fully annotated, multi-sequence dataset, we demonstrated that hetero-modal learning effectively overcomes the rigid data requirements of traditional modality specialist models. Rather than discarding incomplete scans, this approach maximises the utility of all available data. We showed this to be effective in data-constrained scenarios, with hetero-modal models generally outperforming modality specialist baselines.

Beyond accommodating missing data during training, our findings highlight the adaptability of hetero-modal networks at inference. When trained on disjointed datasets sharing only a single anchor modality, these models successfully generalised to novel modality combinations, maintaining robust predictive performance even when the highly informative FLAIR sequence was unavailable.

Additionally, we address a major hurdle in clinical deployment: silent data degradation. By introducing the Multimodal Attention Router (MMAR) block alongside a complimentary corruption augmentation strategy, we showed that this proof-of-concept method can dynamically adapt to unflagged, catastrophically corrupted modalities. Through spatially adaptive feature routing, the MMAR-based model learns to treat severely degraded regions in scans as missing, preserving segmentation accuracy in the presence of extreme noise, bias fields, and false-present modalities. 

Ultimately, these strategies offer a robust and practical pathway for developing and deploying medical image segmentation tools in imperfect, real-world clinical environments.

\paragraph{Limitations}
We acknowledge several limitations in this study. 
First, it is likely that our comparative results are highly sensitive to the total volume and modality distribution of the training data; we urge caution when extrapolating our results to other data distributions.
Second, due to the number of individual training runs required and limited computational resources, our evaluations are centred on a single random training, validation, and test split, and each model was trained only once. While we attempted to mitigate the impact of these factors by performing statistical significance tests, we must highlight this limitation.
Third, our evaluation relies solely on the Dice similarity coefficient (DSC). Although DSC has known limitations, it remains the standard segmentation metric and correlates strongly with other measures. Since our primary objective was the relative comparison of methods rather than establishing absolute task performance, limiting our evaluation to DSC allowed us to reduce the complexity of the results while preserving statistical power.
Fourth, due to a lack of truly corrupted scans in our dataset, test time corruption transformations were required for evaluating the models that were trained for corruption-robustness, using corruption augmentation. To mitigate this bias, we applied different transformation parameters during training and testing augmentations.
Finally, because our primary motivation was to evaluate the viability of hetero-modal learning in data-starved clinical scenarios, we opted for foundational, proven baseline methods. However, an abundance of complex architectures and training techniques have been proposed for imbalanced hetero-modal learning---typically evaluated on the BraTS brain tumour dataset \citep{menze2014multimodal}. We leave a systematic benchmarking of these methods applied to WMH and ISL segmentation for future work.

\section*{Acknowledgements}

J.P. is funded by Medical Research Scotland [ref. PHD-50441-2021] and Canon Medical Research Europe. Funding from Row Fogo Charitable Trust (Ref No: AD.ROW4.35. BRO-D.FID3668413), and the UK Medical Research Council (UK Dementia Research Institute at the University of Edinburgh, award number UK DRI-4002; G0700704/84698) are also gratefully acknowledged. M.O.B. gratefully acknowledges funding from: EPSRC grant no. EP/X025705/1; British Heart Foundation and The Alan Turing Institute Cardiovascular Data Science Award (C-10180357); the SCONe projects funded by Chief Scientist Office, Edinburgh \& Lothians Health Foundation, Sight Scotland, the Royal College of Surgeons of Edinburgh, the RS Macdonald Charitable Trust, and Fight For Sight.


The original study which produced the data used in this paper was support by: the UK Dementia Research Institute which receives its funding from DRI Ltd, funded by the UK MRC, Alzheimer’s Society and Alzheimer’s Research UK; the Fondation Leducq Network for the Study of Perivascular Spaces in Small Vessel Disease (16 CVD 05); Stroke Association ``Small Vessel Disease-Spotlight on Symptoms (SVD-SOS)''(SAPG 19\textbackslash100068); The Row Fogo Charitable Trust Centre for Research into Aging and the Brain; Stroke Association Garfield Weston Foundation Senior Clinical Lectureship (FND) (TSALECT 2015/04); NHS Research Scotland (FND); Stroke Association Post-Doctoral Fellowship (SW) (SAPDF 18/100026); British Heart Foundation Edinburgh Centre for Research Excellence (RE/18/5/34216); NHS Lothian Research and Development Office (MJT); European Union Horizon 2020, PHC-03–15, project No666881, ``SVDs@Target'' (MS,GB); Chief Scientist Office of Scotland Clinical Academic Fellowship (UC) (CAF/18/08); Stroke Association Princess Margaret Research Development Fellowship (UC) (2018); Medical Research Scotland studentship (AM) (PhD-1165–2017); College of Medicine and Veterinary Medicine, University of Edinburgh scholarship, as part of the Wellcome-funded Translational Neuroscience PhD programme (OH); MRC Doctoral Training Programme in Precision Medicine (CM) (MR/R01566X/1). The Research MR scanners are supported by the Scottish Funding Council through the Scottish Imaging Network, A Platform for Scientific Excellence (SINAPSE) Collaboration; the 3T scanner is funded by the Wellcome Trust (104916/Z/14/Z), Dunhill Trust (R380R/1114), Edinburgh and Lothians Health Foundation (2012/17), Muir Maxwell Research Fund and the University of Edinburgh.

\bibliographystyle{elsarticle-harv} 
\bibliography{REFERENCES}

@article{duering2023neuroimaging,
  title={Neuroimaging standards for research into small vessel disease—advances since 2013},
  author={Duering, Marco and Biessels, Geert Jan and Brodtmann, Amy and Chen, Christopher and Cordonnier, Charlotte and de Leeuw, Frank-Erik and Debette, St{\'e}phanie and Frayne, Richard and Jouvent, Eric and Rost, Natalia S and others},
  journal={The Lancet Neurology},
  volume={22},
  number={7},
  pages={602--618},
  year={2023},
  publisher={Elsevier}
}

@article{edlow2017diagnosis,
  title={Diagnosis of DWI-negative acute ischemic stroke: a meta-analysis},
  author={Edlow, Brian L and Hurwitz, Shelley and Edlow, Jonathan A},
  journal={Neurology},
  volume={89},
  number={3},
  pages={256--262},
  year={2017},
  publisher={Lippincott Williams \& Wilkins Hagerstown, MD}
}

@article{phitidis2026comparative,
  title={Comparative evaluation of training strategies using partially labelled datasets for segmentation of white matter hyperintensities and stroke lesions in FLAIR MRI},
  author={Phitidis, Jesse and Smithard, Alison Q and Whiteley, William N and Wardlaw, Joanna M and Bernabeu, Miguel O and Hern{\'a}ndez, Maria Vald{\'e}s},
  journal={arXiv preprint arXiv:2601.20503},
  year={2026}
}

@article{isensee2021nnu,
  title={nnU-Net: a self-configuring method for deep learning-based biomedical image segmentation},
  author={Isensee, Fabian and Jaeger, Paul F and Kohl, Simon AA and Petersen, Jens and Maier-Hein, Klaus H},
  journal={Nature methods},
  volume={18},
  number={2},
  pages={203--211},
  year={2021},
  publisher={Nature Publishing Group US New York}
}

@article{menze2014multimodal,
  title={The multimodal brain tumor image segmentation benchmark (BRATS)},
  author={Menze, Bjoern H and Jakab, Andras and Bauer, Stefan and Kalpathy-Cramer, Jayashree and Farahani, Keyvan and Kirby, Justin and Burren, Yuliya and Porz, Nicole and Slotboom, Johannes and Wiest, Roland and others},
  journal={IEEE transactions on medical imaging},
  volume={34},
  number={10},
  pages={1993--2024},
  year={2014},
  publisher={IEEE}
}

@article{perez2021torchio,
  title={TorchIO: a Python library for efficient loading, preprocessing, augmentation and patch-based sampling of medical images in deep learning},
  author={P{\'e}rez-Garc{\'\i}a, Fernando and Sparks, Rachel and Ourselin, S{\'e}bastien},
  journal={Computer methods and programs in biomedicine},
  volume={208},
  pages={106236},
  year={2021},
  publisher={Elsevier}
}

@article{cardoso2022monai,
  title={Monai: An open-source framework for deep learning in healthcare},
  author={Cardoso, M Jorge and Li, Wenqi and Brown, Richard and Ma, Nic and Kerfoot, Eric and Wang, Yiheng and Murrey, Benjamin and Myronenko, Andriy and Zhao, Can and Yang, Dong and others},
  journal={arXiv preprint arXiv:2211.02701},
  year={2022}
}

@article{clancy2021rationale,
  title={Rationale and design of a longitudinal study of cerebral small vessel diseases, clinical and imaging outcomes in patients presenting with mild ischaemic stroke: Mild Stroke Study 3},
  author={Clancy, Una and Garcia, Daniela Jaime and Stringer, Michael S and Thrippleton, Michael J and Vald{\'e}s-Hern{\'a}ndez, Maria C and Wiseman, Stewart and Hamilton, Olivia KL and Chappell, Francesca M and Brown, Rosalind and Blair, Gordon W and others},
  journal={European stroke journal},
  volume={6},
  number={1},
  pages={81--88},
  year={2021},
  publisher={SAGE Publications Sage UK: London, England}
}

@inproceedings{havaei2016hemis,
  title={Hemis: Hetero-modal image segmentation},
  author={Havaei, Mohammad and Guizard, Nicolas and Chapados, Nicolas and Bengio, Yoshua},
  booktitle={International conference on medical image computing and computer-assisted intervention},
  pages={469--477},
  year={2016},
  organization={Springer}
}

@inproceedings{ronneberger2015u,
  title={U-net: Convolutional networks for biomedical image segmentation},
  author={Ronneberger, Olaf and Fischer, Philipp and Brox, Thomas},
  booktitle={International Conference on Medical image computing and computer-assisted intervention},
  pages={234--241},
  year={2015},
  organization={Springer}
}

@inproceedings{dorent2019hetero,
  title={Hetero-modal variational encoder-decoder for joint modality completion and segmentation},
  author={Dorent, Reuben and Joutard, Samuel and Modat, Marc and Ourselin, S{\'e}bastien and Vercauteren, Tom},
  booktitle={International Conference on Medical Image Computing and Computer-Assisted Intervention},
  pages={74--82},
  year={2019},
  organization={Springer}
}

@inproceedings{zhang2022mmformer,
  title={mmformer: Multimodal medical transformer for incomplete multimodal learning of brain tumor segmentation},
  author={Zhang, Yao and He, Nanjun and Yang, Jiawei and Li, Yuexiang and Wei, Dong and Huang, Yawen and Zhang, Yang and He, Zhiqiang and Zheng, Yefeng},
  booktitle={International conference on medical image computing and computer-assisted intervention},
  pages={107--117},
  year={2022},
  organization={Springer}
}

@inproceedings{liu2023sfusion,
  title={SFusion: Self-attention based n-to-one multimodal fusion block},
  author={Liu, Zecheng and Wei, Jia and Li, Rui and Zhou, Jianlong},
  booktitle={International conference on medical image computing and computer-assisted intervention},
  pages={159--169},
  year={2023},
  organization={Springer}
}

@inproceedings{zhu2025bridging,
  title={Bridging the gap in missing modalities: Leveraging knowledge distillation and style matching for brain tumor segmentation},
  author={Zhu, Shenghao and Chen, Yifei and Chen, Weihong and Wang, Yuanhan and Liu, Chang and Jiang, Shuo and Qin, Feiwei and Wang, Changmiao},
  booktitle={International Conference on Medical Image Computing and Computer-Assisted Intervention},
  pages={95--106},
  year={2025},
  organization={Springer}
}

@article{lan2025dmaf,
  title={DMAF-Net: An Effective Modality Rebalancing Framework for Incomplete Multi-Modal Medical Image Segmentation},
  author={Lan, Libin and Li, Hongxing and Xia, Zunhui and Zhang, Yudong},
  journal={arXiv preprint arXiv:2506.11691},
  year={2025}
}

@inproceedings{shi2024passion,
  title={Passion: Towards effective incomplete multi-modal medical image segmentation with imbalanced missing rates},
  author={Shi, Junjie and Shang, Caozhi and Sun, Zhaobin and Yu, Li and Yang, Xin and Yan, Zengqiang},
  booktitle={Proceedings of the 32nd ACM International Conference on Multimedia},
  pages={456--465},
  year={2024}
}

@article{li2023myops,
  title={MyoPS: A benchmark of myocardial pathology segmentation combining three-sequence cardiac magnetic resonance images},
  author={Li, Lei and Wu, Fuping and Wang, Sihan and Luo, Xinzhe and Mart{\'\i}n-Isla, Carlos and Zhai, Shuwei and Zhang, Jianpeng and Liu, Yanfei and Zhang, Zhen and Ankenbrand, Markus J and others},
  journal={Medical Image Analysis},
  volume={87},
  pages={102808},
  year={2023},
  publisher={Elsevier}
}

@inproceedings{wagner2025feasibility,
  title={Feasibility of federated learning from client databases with different brain diseases and mri modalities},
  author={Wagner, Felix and Xu, Wentian and Saha, Pramit and Liang, Ziyun and Whitehouse, Daniel and Menon, David and Newcombe, Virginia and Voets, Natalie and Noble, J Alison and Kamnitsas, Konstantinos},
  booktitle={2025 IEEE/CVF Winter Conference on Applications of Computer Vision (WACV)},
  pages={357--367},
  year={2025},
  organization={IEEE}
}

@article{lecun1989backpropagation,
  title={Backpropagation applied to handwritten zip code recognition},
  author={LeCun, Yann and Boser, Bernhard and Denker, John S and Henderson, Donnie and Howard, Richard E and Hubbard, Wayne and Jackel, Lawrence D},
  journal={Neural computation},
  volume={1},
  number={4},
  pages={541--551},
  year={1989},
  publisher={MIT Press}
}

@article{vaswani2017attention,
  title={Attention is all you need},
  author={Vaswani, Ashish and Shazeer, Noam and Parmar, Niki and Uszkoreit, Jakob and Jones, Llion and Gomez, Aidan N and Kaiser, {\L}ukasz and Polosukhin, Illia},
  journal={Advances in neural information processing systems},
  volume={30},
  year={2017}
}

@article{kuijf2019standardized,
  title={Standardized assessment of automatic segmentation of white matter hyperintensities and results of the WMH segmentation challenge},
  author={Kuijf, Hugo J and Biesbroek, J Matthijs and De Bresser, Jeroen and Heinen, Rutger and Andermatt, Simon and Bento, Mariana and Berseth, Matt and Belyaev, Mikhail and Cardoso, M Jorge and Casamitjana, Adria and others},
  journal={IEEE transactions on medical imaging},
  volume={38},
  number={11},
  pages={2556--2568},
  year={2019},
  publisher={IEEE}
}

@article{liew2022large,
  title={A large, curated, open-source stroke neuroimaging dataset to improve lesion segmentation algorithms},
  author={Liew, Sook-Lei and Lo, Bethany P and Donnelly, Miranda R and Zavaliangos-Petropulu, Artemis and Jeong, Jessica N and Barisano, Giuseppe and Hutton, Alexandre and Simon, Julia P and Juliano, Julia M and Suri, Anisha and others},
  journal={Scientific data},
  volume={9},
  number={1},
  pages={320},
  year={2022},
  publisher={Nature Publishing Group UK London}
}

@article{hernandez2022isles,
  title={ISLES 2022: A multi-center magnetic resonance imaging stroke lesion segmentation dataset},
  author={Hernandez Petzsche, Moritz R and De La Rosa, Ezequiel and Hanning, Uta and Wiest, Roland and Valenzuela, Waldo and Reyes, Mauricio and Meyer, Maria and Liew, Sook-Lei and Kofler, Florian and Ezhov, Ivan and others},
  journal={Scientific data},
  volume={9},
  number={1},
  pages={762},
  year={2022},
  publisher={Nature Publishing Group UK London}
}

@article{hoopes2022synthstrip,
  title={SynthStrip: skull-stripping for any brain image},
  author={Hoopes, Andrew and Mora, Jocelyn S and Dalca, Adrian V and Fischl, Bruce and Hoffmann, Malte},
  journal={NeuroImage},
  volume={260},
  pages={119474},
  year={2022},
  publisher={Elsevier}
}

@article{tustison2010n4itk,
  title={N4ITK: improved N3 bias correction},
  author={Tustison, Nicholas J and Avants, Brian B and Cook, Philip A and Zheng, Yuanjie and Egan, Alexander and Yushkevich, Paul A and Gee, James C},
  journal={IEEE transactions on medical imaging},
  volume={29},
  number={6},
  pages={1310--1320},
  year={2010},
  publisher={IEEE}
}

@article{valdes2023step,
  title={Step-by-step pipeline for segmenting enlarged perivascular spaces from 3D T2-weighted MRI, 2018-2023 [software]},
  author={Vald{\'e}s Hern{\'a}ndez, MDC and Ballerini, L and Glatz, A and Aribisala, BS and Bastin, ME and Dickie, DA and Duarte Coello, R and Munoz Maniega, S and Wardlaw, JM},
  journal={University of Edinburgh. College of Medicine and Veterinary Medicine. Centre for Clinical Brain Sciences. DOI: https://doi. org/10.7488/ds/7486},
  year={2023}
}

\end{document}